\documentclass[prb,aps,twocolumn,10pt,superscriptaddress,longbibliography,nofootinbib]{revtex4-1}
\usepackage{amsmath}
\usepackage{amsfonts}
\usepackage{graphicx}
\usepackage{bm}
\usepackage{bbold}
\usepackage{color}
\usepackage{braket}
\usepackage{hyperref}
 \hypersetup{
     colorlinks=true,
     linkcolor=blue,
     filecolor=blue,
     citecolor = magenta,      
     urlcolor=red,
     }

\usepackage{xcolor}

\usepackage{letltxmacro}
\LetLtxMacro{\ORIGselectlanguage}{\selectlanguage}
\makeatletter
\DeclareRobustCommand{\selectlanguage}[1]{%
  \@ifundefined{alias@\string#1}
    {\ORIGselectlanguage{#1}}
    {\begingroup\edef\x{\endgroup
       \noexpand\ORIGselectlanguage{\@nameuse{alias@#1}}}\x}%
}
\newcommand{\definelanguagealias}[2]{%
  \@namedef{alias@#1}{#2}%
}

\makeatother

\definelanguagealias{en}{english}
\definelanguagealias{EN}{english}
\definelanguagealias{eng}{english}
\definelanguagealias{de}{ngerman}

\begin{document}

\abovedisplayskip=6pt
\abovedisplayshortskip=6pt
\belowdisplayskip=6pt
\belowdisplayshortskip=6pt

\title{Quantized quasinormal mode description of non-linear cavity QED effects from coupled resonators with a Fano-like resonance}

\author{Sebastian Franke}
\email{sebastian.franke@tu-berlin.de}
\affiliation{Technische Universit\"at Berlin, Institut f\"ur Theoretische Physik,
Nichtlineare Optik und Quantenelektronik, Hardenbergstra{\ss}e 36, 10623 Berlin, Germany}
 \author{Marten Richter}
\affiliation{Technische Universit\"at Berlin, Institut f\"ur Theoretische Physik,
 Nichtlineare Optik und Quantenelektronik, Hardenbergstra{\ss}e 36, 10623 Berlin, Germany}
 \author{Juanjuan Ren}
 \affiliation{\hspace{0pt}Department of Physics, Engineering Physics, and Astronomy, Queen's University, Kingston, Ontario K7L 3N6, Canada\hspace{0pt}}
   \author{Andreas Knorr}
 \affiliation{Technische Universit\"at Berlin, Institut f\"ur Theoretische Physik,
 Nichtlineare Optik und Quantenelektronik, Hardenbergstra{\ss}e 36, 10623 Berlin, Germany}
 \author{Stephen Hughes}
\affiliation{\hspace{0pt}Department of Physics, Engineering Physics, and Astronomy, Queen's University, Kingston, Ontario K7L 3N6, Canada\hspace{0pt}}

\date{\today}

\begin{abstract}
    We employ a recently developed quantization scheme for quasinormal modes (QNMs) to
    study a 
    nonperturbative  open cavity-QED system  
    consisting of a hybrid metal-dielectric resonator coupled to a quantum emitter.
 This hybrid cavity system allows one to explore the complex coupling between a low $Q$ (quality factor) resonance and a high $Q$ resonance, manifesting in a striking Fano resonance, an effect that is not captured by
 traditional quantization schemes using normal modes or a Jaynes-Cummings (JC) type model.
 The QNM quantization approach rigorously includes dissipative coupling between the QNMs, and  is supplemented with generalized input-output relations for the output electric field operator for multiple modes in the system, and correlation functions outside the system. 
 The role of the dissipation-induced mode coupling is explored in the strong coupling regime between the  photons
 and emitter beyond the first rung of the  JC dressed-state ladder. Important differences in the quantum master equation and input-output relations between the QNM quantum model and phenomenological dissipative JC models are found.
 In a second step, numerical results for the Fock distributions and system as well as output correlation functions obtained from the quantized QNM model for the hybrid structure are compared with results from a phenomenological approach.
 We 
 demonstrate explicitly how the quantized QNM
 model manifests in multiphoton quantum correlations beyond what is predicted by the usual JC models. 
\end{abstract}

\maketitle

\section{Introduction\label{Sec: Intro}}

Quantum emitters coupled to photons/plasmons in dissipative nanostructures, such as micropillars~\cite{micropillars,micropillars2,Reithmaier_Nature_432_197_2004}, photonic crystal cavities~\cite{Yoshie_Nature_432_200_2004,sh2007}, dielectric microdiscs~\cite{cao2015dielectric} or metallic nanoparticles~\cite{nanogold2,Akselrod2016,david,NanoMarten,theuerholz2013influence}, constitute an important field in quantum optics and quantum plasmonics. New
classes of nanophotonics structures exhibit enhanced light-matter coupling suitable for studying a range of
interesting phenomena and applications, such as non-classical light generation~\cite{faraon2008coherent,brooks2012non}, spasing~\cite{spaser,PhysRevLett.Spaser,warnakula2019cavity} and quantum information processing~\cite{quantinfo,loss1998quantum}. 
While the dielectric cavities have high quality  factors with state-of-the-art values~\cite{schneider2016quantum} around $Q\sim 10^5$, metallic cavities are significantly more lossy (typically $Q\sim 10$) due to Ohmic heating, but still provide comparable light-matter coupling regimes thanks to the strong local field confinement below the diffraction limit~\cite{maier2003local}. Other recent important lossy structures are hybrid metal-dielectric resonators, made from metals and dielectric resonators~\cite{Barth2010,KamandarDezfouli2017,dezfouli2019molecular,doeleman2016antenna}, which combine the attributes of both resonator types
and exhibit Fano interference effects between both systems.

For the theoretical description of quantum light-matter interactions, bound photon states in such systems are often treated using
``normal modes'' with real eigenfrequencies, typically using Jaynes-Cummings (JC) type models~\cite{Jaynes_ProcIEEE_51_89_1963,carmichael2009statistical}.
Additionally, since these structures are lossy, dissipation is usually introduced into the model by phenomenologically adding decay rates for the subsystems, typically for the optical modes and  quantum emitters~\cite{PhysRevA.82.043845}.
In contrast to this phenomenological approach, so-called quasinormal modes~\cite{Lai,LeungSP1D,Leung3,2ndquant2,kristensen2019modeling} (QNMs) with  complex eigenfrequencies (including loss in the imaginary part), intrinsically  describe the open/lossy system  by solving  Maxwell equations
with open boundary conditions. This approach allows one to determine useful cavity properties such as 
the radiative beta factors (quantum efficiencies), quality factors, and effective mode volumes~\cite{muljarovPert,KristensenHughes,SauvanNorm,NormKristHughes,Lalanne_review,carlson2019dissipative}.
With continued developments in computational electromagnetics, 
the numerical solution of the Helmholtz equation to obtain QNMs is becoming better unraveled and common today \cite{KristensenHughes,Lalanne_review}, but its subsequent quantization using QNMs to retrieve  well-known model systems, e.g., a microscopically defined JC model is still highly nontrivial.

The lossy and non-Hermitian character of these open systems prevents the use of a canonical quantization procedure for the discrete modes of interest (cf.~Refs.~\onlinecite{LeungSP1D, 2ndquanho, Severini}). Recently, a general quantization scheme for three-dimensional absorptive and lossy media using QNMs as the basis for the field expansion was presented~\cite{PhysRevLett.122.213901}, based on a Green function quantization approach~\cite{Dung, grunwel}. It was demonstrated that, for more than one QNM, a coupling between the QNMs is induced by the dissipation through proper quantization in the dissipative system. 
The off-diagonal coupling is especially interesting for mode interference effects in the above mentioned hybrid structures, which is in the semi-classical model a consequence of the complex-valued QNMs\cite{KamandarDezfouli2017}, and leads to highly non-Lorentzian line shapes. In fact, 
Franke {\it et al.}~\cite{PhysRevLett.122.213901} demonstrated, in the single-photon limit (weak excitation), that such an interference effect  can only be reproduced through the off-diagonal QNM coupling, when starting from a quantized mode approach. This was further confirmed in Ref.~\onlinecite{denning2019quantum} for a Fano cavity, by a independent method and calculation based on introducing a phenomenological mode coupling in a two-mode master equation.  

Recently, the approach from Ref.~\onlinecite{PhysRevLett.122.213901} was  
applied to 
accurately  describe single-photon emission in a single-mode metal resonator~\cite{hughes2019theory} and was also used to model the photonic mode quantization for molecular optomechanics in a hybrid metal-dielectric resonator~\cite{dezfouli2019molecular}. In the first case, a cavity output field expression for the single-QNM was derived, which is the basis to determine correlation functions and light statistics of the resonator-emitter system, important to simulate experimental situation, such as the Hong-Ou-Mandel~\cite{hong1987measurement} or Hanbury-Brown-Twiss~\cite{brown1957interferometry} setups. Importantly, the QNM quantization scheme allows one to distinguish between radiative and non-radiative decay processes, which both enter naturally into the formalism through the same calculated QNMs. This separation is essential to describe a realistic input-output formalism in cases of absorptive lossy structures, where the output is usually treated in the same way as for systems with only radiative losses~\cite{koenderink2010use, koenderink2017single, ren2017evanescent}.  While in the single-mode case, the results from a more phenomenological dissipative JC model is modified by a 
loss-induced prefactor, which separates between radiative and non-radiative decay~\cite{hughes2019theory}, there are additional changes in the multi-mode case due to off-diagonal mode interaction, which may also effect the output coupling. This may also effect the behavior of multi-mode systems in higher rungs of the JC ladder, e.g., the change of Poissionian to sub-Possionian light when coupling a resonator-emitter system to another resonator, which is often described with two uncoupled modes~\cite{majumdar2012loss, zhang2014optimal}.

\begin{figure*}[ht]
	\centering
	\includegraphics[width=2\columnwidth,trim={0 0 0 0},clip]{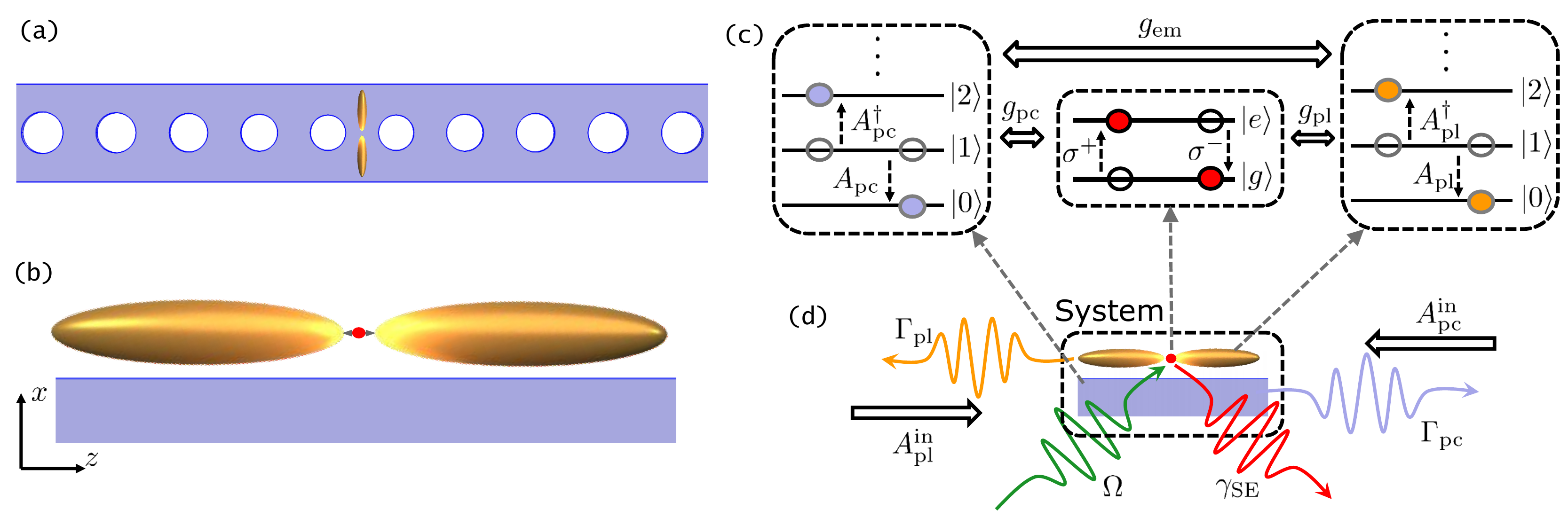}
 	\caption{(a) Top view on the metal-dielectric hybrid structure, consisting of a metallic dimer on top of a photonic crystal beam. (b) Side view on the metallic dimer, supporting a quantum dipole ($z$-polarized) in the middle of the gap. (c) Schematic of the quantum subsystems: The photonic crystal-like mode (violet) is coupled to the plasmonic-like cavity mode (yellow) and the TLS (red) is coupled to both symmetrized QNMs. d) Excitation and input-output scheme. The TLS is driven by a cw-pump field with strength $\Omega_{\rm L}$ and decays with a rate $\gamma_{\rm SE}$, while the QNM subsystem is characterized by the decay rates $\Gamma_{\rm pl}$ and $\Gamma_{\rm pc}$, naturally entering the model through the input fields $A^{\rm in}_{\rm pl}$ and $A^{\rm in}_{\rm pc}$, respectively (see text).
	}\label{fig: Scheme}
\end{figure*}

In this paper, we study the nonlinear multiphoton cavity-QED effects of a hybrid cavity structure containing a single quantum emitter, depicted in Fig.~\ref{fig: Scheme}(a,b), which consists of a metal ellipsoid dimer on top of a photonic crystal beam.  
In contrast to the hybrid structure used in the Ref.~\onlinecite{PhysRevLett.122.213901}, the resonator-TLS system in Fig.~\ref{fig: Scheme}(a,b) is in the strong coupling regime, where the bad cavity approximation is no longer valid.
Therefore, processes on higher rungs, i.e., many photon effects, in the anharmonic JC-model are more accessible and we analyse the effects of the inter-mode coupling in this multiphoton regime. To calculate measurable quantities, such as the second-order correlation function, an expression for the electric field operator outside the resonator for multiple QNMs is derived.
A second main objective of this work is to compare the results for the Hamiltonian and Liouvillian of the quantized QNM model 
with a phenomenological dissipative two-mode JC model in the few photon limit. The phenomenological model assumes two uncoupled bosonic modes. It will be shown, that the off-diagonal coupling present in the quantization of the QNMs will induce drastic changes in the master equation and density matrix simulations for higher rungs of the JC ladder, which will further underline the importance of a quantized QNM model.

The rest of our paper is organized as follows: In Section~\ref{Sec: Theor}, we present the theoretical framework for this paper. First, we summarize the phenomenological Green function quantization approach for general spatial-inhomogeneous and absorptive media; this includes a coupling between the medium-assisted electromagnetic field and an emitter. Second, the definition and main aspects of the QNM approach will be revisited and clarified. Third, we will briefly recapitulate the QNM quantization from Ref.~\onlinecite{PhysRevLett.122.213901}. Then, the equations of motion for the QNM operators and input-output relation for multiple QNMs in the system are derived, and, based on that derivation, the QNM master equation with an additional external pump term is presented. Last, we derive the multi-mode output electric field operator to express the output correlation function in terms of QNM operators.

In Section~\ref{Sec: Applications}, we present three-dimensional numerical results for a metal-dielectric hybrid structure (see Fig.~\ref{fig: Scheme}(a)) using the two-mode phenomenological dissipative JC model and the rigorous QNM quantum master equation. First, we show results for the Purcell factor of the quantum emitter. Second, we analyse the system properties, including the Fock distributions and populations.  
Third, we  use the derived expressions of the output electric field operators from Section~\ref{Sec: Theor} and show results for the output photon correlation functions. 
In Section~\ref{Sec: Conlusions}, we will summarize our results and give an outlook to future applications of the theory. 
We complement the main part of this work with six appendices, that contain a more throughout derivation of the QNM input-output relations, details of the QNM parameters of the hybrid structure, a more detailed derivation of the photon correlation functions, as well as discussions about the light-matter coupling regimes of the hybrid structure, the response of the hybrid cavity to the external driving, and the treatment of the frequency integrals in the case of a few QNM expansion. 

\section{Theory\label{Sec: Theor}}
\subsection{Green function quantization approach\label{Subsec: GFquant}}
We start with the Hamiltonian $H=H_{\rm a}+H_{\rm B}+H_{\rm I}$, where $H_{\rm a}$ describes a two level system (TLS) with a transition frequency $\omega_{\rm a}$, interacting with a electromagnetic field in an absorptive and spatial-inhomogeneous  media using the quantization scheme from Refs.~\onlinecite{grunwel,Dung,vogel2006}. The term $H_{\rm B}$ contains the energy of the medium-assisted electromagnetic field, and $H_{\rm I}$ is a dipole-field interaction Hamiltonian (in the rotating wave approximation). The total Hamiltonian  thus contains the following contributions:
\begin{align}
    H_{\rm a} &= \hbar\omega_{\rm a} \sigma^+\sigma^-,\label{eq: Ha}\\
    H_{\rm B} &= \hbar\int{\rm d}\mathbf{r}\int_0^\infty {\rm d}\omega~ \omega ~\mathbf{b}^\dagger(\mathbf{r},\omega)\cdot \mathbf{b}(\mathbf{r},\omega),\label{eq: HB}\\
    H_{\rm I}&=-\sigma^+\int_0^{\infty}{\rm d}\omega \mathbf{d}_{\rm a}\cdot\left\{\hat{\mathbf{E}}(\mathbf{r}_{\rm a},\omega)+\mathbf{E}_{\rm L}(\mathbf{r}_{\rm a},\omega)\right\}-{\rm H.a.},\label{eq: HI}
\end{align}
where $\mathbf{b}^{(\dagger)}(\mathbf{r},\omega)$ are annihilation (creation) operators acting on the combined Hilbert space of the dissipative medium and the electromagnetic field degrees of freedom, represented by the spatial index $\mathbf{r}$ and frequency index $\omega$. The variable $\omega$ must be regarded as continuous mode index, rather then a (temporal) Fourier variable. In fact, in the Heisenberg picture the fundamental operators are $\mathbf{b}^{(\dagger)}(\mathbf{r},\omega,t)$, and the time evolution is governed by the Heisenberg equation of motion with respect to $H$. We further note, that $\mathbf{b}^{(\dagger)}(\mathbf{r},\omega)$ fulfills canonical commutation relations. 
The terms $\sigma^{\pm}$ denote lowering and raising operators of the TLS, describing a point-like emitter with dipole moment $\mathbf{d}_{\rm a}$ at the position $\mathbf{r}_{\rm a}$. 

The TLS 
interacts with an effective semi-classical excitation field $\mathbf{E}_{\rm L}(\mathbf{r}_{\rm a},\omega)$, which reflects a contribution of an incident laser field at the quantum emitter position, that is enhanced by the scattering structures in the dielectric medium, and a medium-assisted quantized electromagnetic field $\hat{\mathbf{E}}(\mathbf{r}_{\rm a},\omega)$, which obeys the quantized Helmholtz equation: 
\begin{equation}
    \boldsymbol{\nabla}\times\boldsymbol{\nabla}\times\hat{\mathbf{E}}(\mathbf{r},\omega)-k_0^2\epsilon(\mathbf{r},\omega)\hat{\mathbf{E}}(\mathbf{r},\omega)=i\omega\mu_0\, \hat{\mathbf{j}}_{\rm N}(\mathbf{r},\omega)\label{eq: HelmholtzE},
\end{equation}
where $k_0=\omega/c$, and $\epsilon(\mathbf{r},\omega)=\epsilon_R(\mathbf{r},\omega)+i\epsilon_I(\mathbf{r},\omega)$ is the dielectric permittivity.
The 
noise current density $\hat{\mathbf{j}}_{\rm N}(\mathbf{r},\omega)=\omega\sqrt{\hbar\epsilon_0\epsilon_I(\mathbf{r},\omega)/\pi}\,\mathbf{b}(\mathbf{r},\omega)$ counteracts the dissipation, such that the commutation relations between the electromagnetic field operators is spatially preserved for the dissipative materials~\cite{Dung,philbin2010canonical,suttorp2004field} as well as  non-dissipative dielectrics~\cite{drezet2017equivalence,franke2020fluctuation}.
A formal solution of Eq.~\eqref{eq: HelmholtzE} is the source-field expression
\begin{equation}
    \hat{\mathbf{E}}(\mathbf{r},\omega) =i \sqrt{\frac{\hbar}{\pi\epsilon_0}}\int{\rm d}\mathbf{r}'\sqrt{\epsilon_I(\mathbf{r}',\omega)}\mathbf{G}(\mathbf{r},\mathbf{r}',\omega)\cdot\mathbf{b}(\mathbf{r}',\omega),\label{eq: SolE}
\end{equation}
where $\mathbf{G}(\mathbf{r},\mathbf{r}',\omega)$ is the photon Green function, solving the usual Helmholtz equation 
\begin{equation}
    \boldsymbol{\nabla}\times\boldsymbol{\nabla}\times\mathbf{G}(\mathbf{r},\mathbf{r}',\omega)-k_0^2\epsilon(\mathbf{r},\omega)\mathbf{G}(\mathbf{r},\mathbf{r}',\omega)=k_0^2\mathbb{1}\delta(\mathbf{r}-\mathbf{r}')\label{eq: HelmholtzG}, 
\end{equation}
together with suitable  boundary conditions. We emphasize again, that the coupling between the  total electric field and the TLS is assumed to be  below the so-called ultrastrong coupling regime~\cite{frisk_kockum_ultrastrong_2019,RevModPhys.91.025005}, i.e., $|\mathbf{d}_{\rm a}\cdot\hat{\mathbf{E}}^{(+)}(\mathbf{r}_{\rm a})|\ll \hbar\omega_{\rm a}$ (where $\hat{\mathbf{E}}^{(+)}(\mathbf{r}_a)=\int_0^\infty{\rm d}\omega \hat{\mathbf{E}}(\mathbf{r},\omega)+\mathbf{E}_{\rm L}(\mathbf{r}_{\rm a},\omega)$), 
which is consistent with our rotating wave approximation.

\subsection{Quantized quasinormal mode approach\label{Subsec: QNMquant}}
After presenting the quantized Maxwell theory for absorptive systems with a continuous set of modes, we now briefly recapitulate the definition and properties of the QNMs (discrete modes, with complex frequencies), and the Green function expansion. We also discuss the regularized QNMs for expanding the quantized medium-assisted electric field operator (outside the system). 

The QNMs for open systems can be viewed in a similar way as the normal modes for closed systems; 
the QNM vector-valued functions $\tilde{\mathbf{f}}_\mu(\mathbf{r})$ are solutions to the Helmholtz equation 
\begin{equation}
\boldsymbol{\nabla}\times\boldsymbol{\nabla}\times\tilde{\mathbf{f}}_{\mu}(\mathbf{r})-\frac{\tilde{\omega}_\mu^2}{c^2}\epsilon(\mathbf{r},\tilde{\omega}_\mu)\tilde{\mathbf{f}}_{\mu}(\mathbf{r})=0\label{eq: HelmholtzQNM},
\end{equation}
however, with open boundary conditions, e.g., the Silver-M\"uller radiation conditions:~\cite{martin2006multiple} 
\begin{equation}
    \frac{\mathbf{r}}{r}\times\boldsymbol{\nabla}\times\tilde{\mathbf{f}}_{\mu}(\mathbf{r})\longrightarrow-in_{\rm B}\frac{\tilde{\omega}_{\mu}}{c}\tilde{\mathbf{f}}_{\mu}(\mathbf{r}),\label{eq: SilverMuller}
\end{equation}
for $|\mathbf{r}|\rightarrow \infty$. Here, $\epsilon(\mathbf{r},\tilde{\omega}_{\mu})$ is the analytical continuation of the permittivity into the complex plane, and we assume in addition to the lossy media $\epsilon(\mathbf{r},\omega)$ a background region with homogeneous refractive index $n_{\rm B}$, in which the resonator structure is embedded. 

As a consequence of the open boundary conditions, the QNM eigenfrequencies  $\tilde{\omega}_\mu=\omega_{\mu}-i\gamma_{\mu}$ are complex numbers with a negative imaginary part, i.e., $\gamma_{\mu}>0$. In combination with the fact, that the corresponding (classical) QNM electric fields are harmonic solutions of the wave equation, i.e., $\tilde{\mathbf{E}}_\mu(\mathbf{r},t)\propto \tilde{\mathbf{f}}_{\mu}(\mathbf{r})\exp(-i\tilde{\omega}_\mu t)$,  
this leads to a lossy character of these modes. Moreover,  
the QNM eigenfunctions behave in the far field as $\tilde{\mathbf{f}}_\mu(\mathbf{r})\propto \exp(in_{\rm B}\tilde{\omega}_\mu |\mathbf{r}|/c)$, and because of the complex eigenvalues (with negative imaginary part) they spatially diverge for far field positions. This leads to an nonphysical behavior~\footnote{It is nonphysical in the sense of representing the total field, which must be finite, but it is still the mathematically correct mode.}
of the QNMs outside of the resonator, but is, in fact, a property of any solution to a Helmholtz equation for a lossy geometry with open boundary conditions.
In spite of this spatial divergence, when properly normalized~\cite{MDR1,muljarovPert,SauvanNorm,normaliz}, the QNMs can be used to expand the photonic Green function in the form~\cite{MDR1,muljarovPert,KristensenHughes,Doost_PRA_87_043827_2013,SauvanNorm}
\begin{equation}
    \mathbf{G}_{\rm ff}(\mathbf{r},\mathbf{r}',\omega)=\sum_{\mu}A_\mu(\omega)\tilde{\mathbf{f}}_{\mu}(\mathbf{r})\tilde{\mathbf{f}}_{\mu}(\mathbf{r}'),\label{eq: GreenQNM}
\end{equation}
for positions $\mathbf{r},\mathbf{r}'$ nearby (or within) the resonator and where 
\begin{equation} 
    A_\mu(\omega) = \frac{\omega}{2(\tilde{\omega}_\mu - \omega)}\label{eq: Amuform1}
\end{equation}
is the QNM Green function expansion coefficient.

We remark that there exist alternative forms of $A_\mu(\omega)$, which can be obtained by applying a sum rule of the QNMs\cite{MDR1}, i.e., 
\begin{equation}
    \sum_\mu \frac{\tilde{\mathbf{f}}_{\mu}(\mathbf{r})\tilde{\mathbf{f}}_{\mu}(\mathbf{r}')}{\tilde{\omega}_\mu}=0.
\end{equation}
In this way, one can find an equivalent form of $A_\mu(\omega)$ in $\mathbf{G}_{\rm ff}(\mathbf{r},\mathbf{r}',\omega)$, defined as
\begin{equation} 
    \tilde{A}_\mu(\omega) = \frac{\omega^2}{2\tilde{\omega}_\mu(\tilde{\omega}_\mu - \omega)}.
\end{equation}
However, it was shown in Ref.~\onlinecite{kristensen2019modeling}, that, upon using a Riesz projection technique, the latter form seems to lack a pole contribution at $\omega=0$, which can lead to an unphysical behaviour of the Green function, when it is not complemented by  additional contributions. In contrast, the form in Eq.~\eqref{eq: Amuform1} is precisely the expansion coefficient for the total (transverse) Green function, when more general Green functions (including poles at $\omega=0$) are considered. Therefore, in the following, we will use the arguably more general form in Eq.~\eqref{eq: Amuform1}.

For positions, $\mathbf{R}$, outside the resonator region, we replace  $\tilde{\mathbf{f}}_{\mu}(\mathbf{R})$ with a regularized QNM $\tilde{\mathbf{F}}_{\mu}(\mathbf{R},\omega)$ via the Dyson equation~\cite{GeNJP2014,PhysRevLett.122.213901},
\begin{equation}
    \tilde{\mathbf{F}}_{\mu}(\mathbf{R},\omega) = \int_V{\rm d}\mathbf{r} \Delta\epsilon(\mathbf{r},\omega)\mathbf{G}_{\rm B}(\mathbf{R},\mathbf{r},\omega)\cdot\tilde{\mathbf{f}}_\mu(\mathbf{r}),\label{eq: DysonF}
\end{equation}
where $\mathbf{G}_{\rm B}(\mathbf{R},\mathbf{r},\omega)$ is the background Green function, solving the Helmholtz equation~\eqref{eq: HelmholtzG} for $\epsilon(\mathbf{r},\omega)=\epsilon_{\rm B}(\omega)$; and   $\Delta\epsilon(\mathbf{r},\omega)=\epsilon(\mathbf{r},\omega)-\epsilon_{\rm B}(\omega)$ is the permittivity difference, where $\epsilon_{\rm B}(\omega)=n_{\rm B}^2$ is the homogeneous permittivity of the background region. 

Alternatively, as shown in Ref.~\onlinecite{ren2020near}, one can approximate the expression in Eq.~\eqref{eq: DysonF} with a near-field to far-field transformation using the field equivalence principle~\cite{Neartofar_1992}: 
\begin{align}
    \tilde{\mathbf{F}}_{\mu}(\mathbf{R},\omega)=&i\omega\mu_0 \oint_{\mathcal{S}'} {\rm d}A_{\mathbf{s}'} \mathbf{G}_{\rm B}(\mathbf{R},\mathbf{s}',\omega)\cdot \mathbf{J}_{\rm S'}^{\mu} (\mathbf{s}')\nonumber\\
    &-\oint_{\mathcal{S}'} {\rm d}A_{\mathbf{s}'}\left[\boldsymbol{\nabla}\times \mathbf{G}_{\rm B}(\mathbf{R},\mathbf{s}',\omega)\right]\cdot \tilde{\mathbf{M}}_{\rm S'}^{\mu} (\mathbf{s}'), \label{eq: NF2FF_Z}
\end{align}
where the terms 
\begin{align}
\tilde{\mathbf{J}}_{\rm S'}^{\mu}(\mathbf{s}')=\mathbf{\hat{n}}'\times\tilde{\bf{h}}_{\mu}(\mathbf{s}'),\\
\tilde{\mathbf{M}}_{\rm S'}^{\mu}(\mathbf{s}')=-\mathbf{\hat{n}}'\times\tilde{\bf{f}}_{\mu}(\mathbf{s}'),
\end{align}
are the sources on the boundary, and $\tilde{\mathbf{h}}_{\mu}(\mathbf{r}')=\nabla\times\mathbf{\tilde{f}_{\mu}(r')}/(i\tilde{\omega}_{\mu}\mu_{0})$ is the magentic field of the associated QNM $\mu$ and $\mathbf{\hat{n}}'$ is the normal vector on the surface $\mathcal{S}'$. 
In contrast to the regularized QNM in Eq.~\eqref{eq: DysonF}, the expression $\tilde{\mathbf{F}}_\mu(\mathbf{R},\omega)$ obtained from the near field to far field transformation, requires the QNM source quantities on a surface $\mathcal{S}'$ surrounding the resonator region 
 (for details, cf. Ref.~\onlinecite{ren2020near}). 

Using the Green function expansion in Eq.~\eqref{eq: GreenQNM} with the regularized QNMs, 
obtained
either from the Dyson approach or the near field to far field transformation, 
gives an approximated form of the  photonic Green function in terms of QNMs for positions inside and outside the resonator. 
Using 
 Eq.~\eqref{eq: SolE} together with Eq.~\eqref{eq: GreenQNM} and the regularization (either Eq.~\eqref{eq: DysonF} or Eq.~\eqref{eq: NF2FF_Z}), we can 
 formulate the total electric field $\hat{\mathbf{E}}(\mathbf{r})=\int_0^\infty{\rm d}\omega \hat{\mathbf{E}}(\mathbf{r},\omega)+{\rm H.a.}$ as~\cite{PhysRevLett.122.213901}
\begin{equation}
 \hat{\mathbf{E}}(\mathbf{r})=i\sum_{\mu}\sqrt{\frac{\hbar\omega_{\mu} }{2\epsilon_0}}\, \tilde{\mathbf{f}}_{\mu}(\mathbf{r}) \tilde{\alpha}_{\mu} + \text{H.a.},\label{eq:E_multi}
 \end{equation}
 which explicitly expands in terms of the QNMs. Here, $\mathbf{r}$ is the position in the resonator region, and we have introduced the QNM operators:  
 \begin{align}
    \tilde{\alpha}_\mu=\sqrt{\frac{2}{\pi\omega_\mu}}\int_0^\infty {\rm d}\omega A_\mu(\omega)\int{\rm d}\mathbf{r}\sqrt{\epsilon_I(\mathbf{r},\omega)}\tilde{\mathbf{f}}_\mu(\mathbf{r})\cdot\mathbf{b}(\mathbf{r},\omega),\label{eq: aDef}
\end{align}
 where $\tilde{\mathbf{f}}_\mu(\mathbf{r})$ is replaced by the regularized QNM $\tilde{\mathbf{F}}_\mu(\mathbf{r},\omega)$ for positions outside the resonator.
These operators fulfill non-bosonic commutation relations, i.e., $[\tilde{\alpha}_\mu,\tilde{\alpha}_\eta^\dagger]=S_{\mu\eta}$, where $S_{\mu\eta}$ is a dissipation-induced coupling matrix between QNMs $\mu,\eta$ and has the form 
\begin{equation}
S_{\mu\mu '} = \int_0^{\infty}{\rm d}\omega\frac{2A_\mu(\omega)A_{\mu'}^*(\omega)}{\pi\sqrt{\omega_\mu\omega_\mu'}}\left(S^{\rm nrad}_{\mu\mu'}(\omega)+S^{\rm rad}_{\mu\mu'}(\omega)\right)\label{eq: Scommute},
\end{equation}
where 
\begin{equation}
S^{\rm nrad}_{\mu\mu'}(\omega) = \int_{V} {\rm d}\mathbf{r}\epsilon_I(\mathbf{r},\omega)\tilde{\mathbf{f}}_{\mu}(\mathbf{r})\cdot\tilde{\mathbf{f}}_{\mu'}^*(\mathbf{r}),\label{eq: Snrad}
\end{equation}
describes nonradiative loss processes, i.e., Ohmic heating into the lossy medium, and 
\begin{align}
    S^{\rm rad}_{\mu\mu'}(\omega)
    =&\frac{1}{2\omega\epsilon_0}\oint_{\mathcal{S}} {\rm d}A_\mathbf{s}\left[\hat{\mathbf{n}}_{\mathbf{s}}\times\tilde{\mathbf{H}}_{\mu}(\mathbf{s},\omega)\right]\cdot\tilde{\mathbf{F}}_{\mu'}^*(\mathbf{s},\omega)\nonumber\\
    &+{\rm H.c.}(\mu\leftrightarrow\mu')\label{eq: Sradpre}, 
\end{align}
accounts for  the radiative loss,
where $\tilde{\mathbf{H}}_{\mu}(\mathbf{s},\omega)=\boldsymbol{\nabla}\times\tilde{\mathbf{F}}_{\mu}(\mathbf{s},\omega)/(i\omega\mu_0)$ is the regularized QNM magnetic field and $\hat{\mathbf{n}}$ points outwards of $\mathcal{S}$. Importantly, we have radiative loss even if the material
system is not lossy, as explained in more detail in Ref.~\onlinecite{franke2020fluctuation}.

Numerically, we have verified that the radiative part of $S$ has to be chosen in the far field (i.e., at least half a wavelength away from the resonator), in order to get a convergent value of $S^{\rm rad}_{\mu\mu'}$. This is clear, since otherwise one gets near-field evanescent contributions that can even be negative, and these are not associated with far field propagation decay. Therefore, by choosing $\mathcal{S}=\mathcal{S}_{\infty}$ as a far field surface and applying Silver-M\"uller radiation condition, we obtain the form 
\begin{align}
    S^{\rm rad}_{\mu\mu'}(\omega)=\frac{n_{\rm B}c}{\omega}\oint_{\mathcal{S}_\infty} {\rm d}A_\mathbf{s}\tilde{\mathbf{F}}_{\mu}(\mathbf{s},\omega)\cdot\tilde{\mathbf{F}}_{\mu'}^*(\mathbf{s},\omega)\label{eq: Srad}.
\end{align}

After  applying a symmetrization orthogonalization transformation~\cite{PhysRevLett.122.213901}, we can rewrite the electric field operator from Eq.~\eqref{eq:E_multi} in a symmetrized QNM basis,
 \begin{equation}
 \hat{\mathbf{E}}(\mathbf{r})=i\sum_{\mu}\sqrt{\frac{\hbar\omega_{\mu} }{2\epsilon_0}}\, \tilde{\mathbf{f}}_{\mu}^s(\mathbf{r}) a_{\mu} + \text{H.a.},\label{eq:Esymm_multi}
 \end{equation}
 with the symmetrized QNM functions,
  \begin{equation}
 \tilde{\mathbf{f}}^{s}_{\mu}(\mathbf{r})=\sum_{\nu}(\mathbf{S}^{\frac{1}{2}})_{\nu\mu}\sqrt{\omega_{\nu}/\omega_{\mu}}\tilde{\mathbf{f}}_{\nu}(\mathbf{r}),
 \end{equation}
and 
$a_\mu=\sum_{\eta}(\mathbf{S}^{-1/2})_{\mu\eta}\tilde{\alpha}_\eta$
and $a_\mu^\dagger$
are the annihilation and creation operators for the symmetrized QNMs.
As a consequence of the symmetrization~\cite{PhysRevLett.122.213901}, these
operators ($a_{\mu}$ and $a_{\mu}^\dagger$) obey bosonic commutation relations and can be used to describe Fock number states of mixed photon/lossy medium excitations. These photon number states are not eigenstates of the photon Hamilton operator, namely Eq.~\eqref{eq: HB}, since the photon number operators do not commute with the Hamiltonian $H_{\rm B}$.

\subsection{Multi quasinormal-mode
master equation\label{Subsec: QNMmaster}}
Next, we provide a more detailed derivation of QNM master equation, introduced in Ref.~\onlinecite{PhysRevLett.122.213901}, also taking into account the external laser field. Exploiting the general quantized QNM theory developed in the last subsection, we can determine the time evolution of the QNM annihilation operator, $a_\mu$, with respect to the Hamiltonian $H$ in Eqs.~\eqref{eq: Ha}-\eqref{eq: HI}. Using the Heisenberg equations of motion, 
we derive within the Markov approximation (cf. App.~\ref{app: InputOpApprox})
\begin{align}
\dot{a}_\mu{=}&-\frac{i}{\hbar}[a_\mu,H_{\rm sys}]-\sum_\eta\chi^{(-)}_{\mu\eta}a_\eta\nonumber\\
&-\sqrt{2}\sum_\eta\left[\left(\boldsymbol{\chi}^{(-)}\right)^{1/2}\right]_{\mu\eta}a^{\rm in}_{\eta},\label{eq:QLE}
\end{align}
where we have defined 
$H_{\rm sys}=H_{\text{em}}{+}H_{\rm a}{+}H_{\text{em-a}}{+}H_{\rm L}$  as the effective system Hamiltonian in the symmetrized QNM basis, with $H_{\text{em}}=\hbar\chi^{(+)}_{\mu\eta}a_\mu^{\dagger}a_\eta$ and $H_{\text{em-a}}=\hbar \sum_\mu\tilde{g}^s_\mu a_\mu\sigma^+ +{\rm H.a.}$,  and we have
introduced the symmetrized QNM-TLS coupling constant $\tilde{g}^s_\mu=\sqrt{\omega_{\rm \mu}/(2\epsilon_0 \hbar)}\mathbf{d}_{\rm a} {\cdot} \tilde{\mathbf{f}}^{s}_\mu (\mathbf{r}_{\rm a})$ and the coherent mode-coupling term $\chi^{(+)}_{\mu\eta}=(\chi_{\mu\eta} + \chi_{\eta\mu}^*)/2$, where $\chi_{\mu\eta}{=}\sum_{\nu}
(\mathbf{S}^{-\frac{1}{2}})_{\mu\nu}\tilde{\omega}_\nu\big(\mathbf{S}^{\frac{1}{2}})_{\nu\eta}$. We included only terms in rotating wave approximation in $H_{\rm em-a}$, since our calculation will not enter the ultrastrong coupling regime, as discussed in subsection~\ref{Subsec: GFquant}. 
The pumping term, $H_{\rm L}$, accounts for the interaction of the dipole with the external laser field (cf. Eq.~\eqref{eq: HI}).

Equation~\eqref{eq:QLE} has the form of a quantum Langevin equation for coupled harmonic oscillators with two additional terms: a damping contribution $\sum_\eta\chi^{(-)}_{\mu\eta}a_\eta$ associated with the QNM radiative and non-radiative decay matrix $\chi^{(-)}_{\mu\eta}=i(\chi_{\mu\eta} - \chi_{\eta\mu}^*)/2$,  and a noise input operator $a^{\rm in}_\mu=\int_0^{\infty}{\rm d}\omega a^{\rm in}_\mu(\omega)$,
(Eq.~\eqref{eq: aInOpMarkov}), representing a quantum Langevin force 
counteracting the damping ~\cite{QNoise}. Indeed, the presence of $a_{\rm in}$ preserves the equal-time commutation relation $[a_\mu(t),a^{\dagger}_\eta(t)]{=}\delta_{\mu\eta}$.

We next derive the QNM master equation based on the quantum Langevin equation in Eq.~\eqref{eq:QLE}.
We first assume that the incident laser field reflects a cw excitation with a detuning 
with respect to the TLS frequency $\omega_{\rm a}$, i.e., such that the effective classical scattered field, i.e., the incident laser field enhanced by the cavity structure, can be approximated as $\mathbf{E}_{\rm L}(\mathbf{r}_{\rm a},\omega,t)\approx\mathbf{F}_{\rm L}(\mathbf{r}_{\rm a})e^{-i\omega_{\rm L} t}$. We subsequently rewrite the dipole-laser interaction Hamiltonian from $H_{\rm I}$ of Eq.~\eqref{eq: HI} as
\begin{equation}
    H_{\rm L}= \hbar\Omega_{\rm L}\left(  e^{-i\omega_{\rm L} t}\sigma^+ +  e^{i\omega_{\rm L} t}\sigma^-\right),
\end{equation}
where $\hbar\Omega_{\rm L} = -\mathbf{d}_{\rm a}\cdot\mathbf{F}_{\rm L}(\mathbf{r}_{\rm a})$ is the Rabi frequency of the enhanced classical laser field. It should be noted, that although we choose here an effective driving of the quantum emitter, there is an equivalence of cavity pumping and quantum emitter pumping, in the sense that the cavity driving leads to an effective quantum emitter driving via the cavity-exciton interaction~\cite{fischer2018pulsed}.
Furthermore, we treat the input operators as white noise, and
 assume that the corresponding input state is the vacuum state, i.e., there are initially zero quanta in the input states, such that $\langle a^{\rm in}_\mu(t)a^{\rm in\dagger}_\eta(t')\rangle = \delta_{\mu\eta}\delta(t{-}t')$ and all other second order correlation functions vanish~\cite{Gardiner1}. In addition, we assume that the eigenfrequencies of $\tilde{H}_{\rm sys}$ are not degenerate~\cite{Lax,Gardiner1}.

Applying the Ito-Stratonovich calculus~\cite{Gardiner1} to the Heisenberg equation of motions of the symmetrized QNM operators, and using the procedure from Ref.~\onlinecite{PhysRevLett.122.213901}, 
 we obtain the master equation for the symmetrized QNMs and TLS (in a rotating frame with respect to $\omega_{\rm L}$):
\begin{equation}
    \partial_t \rho = -\frac{i}{\hbar}[H_{\rm sys}',\rho]+\mathcal{L}_{\rm em}\rho+\mathcal{L}_{\rm SE}\rho\label{eq: QNMMaster},
\end{equation}
where $H_{\rm sys}'=H_{\rm a}' + H'_{\rm em}+H'_{\rm L}+H_{\rm em-a}$ is the effective system Hamiltonian (described after Eq.~\eqref{eq:QLE}) with $H_{\text{em}}'=\hbar\sum_{\mu\eta}(\chi^{(+)}_{\mu\eta}-\delta_{\mu\eta}\omega_{\rm L})a_\mu^{\dagger}a_\eta$,  
 $H_{\rm a}'= \hbar\Delta_{\rm a}\sigma^+\sigma^-$, and $H'_{\rm L} = \hbar\Omega(\sigma^+ + \sigma^-)$ in the rotating frame with the (laser) detuned TLS frequency $\Delta_{\rm a}=\omega_{\rm a} - \omega_{\rm L}$. We stress again, that due to inter-mode coupling terms, the photon number operators $N_\mu=a_\mu^\dagger a_\mu$ do not commute with the Hamiltonian, as would be usually the case in a Fock space without dissipation.
 
The QNM Lindblad dissipator is derived as 
\begin{equation}
    \mathcal{L}_{\rm em}\rho =\sum_{\mu,\eta} \chi_{\mu\eta}^{(-)}\left[2a_\eta\rho a_\mu^\dagger - a_\mu^\dagger a_\eta \rho - \rho a_\mu^\dagger a_\eta\right],
\end{equation}
which also yields an off-diagonal coupling via the decay matrix $\chi_{\mu\eta}^{(-)}$.
We further added the Lindblad dissipator, 
\begin{equation}
    \mathcal{L}_{\rm SE}\rho= \frac{\gamma_{\rm SE}}{2}\left[2\sigma^-\rho \sigma^+ - \sigma^+\sigma^-\rho - \rho \sigma^+\sigma^-\right]\label{eq: LindSE}, 
\end{equation}
with the (background) spontaneous emission (SE) rate
\begin{align}
    \gamma_{\rm SE} &= \frac{2}{\hbar\epsilon_0}\mathbf{d}_{\rm a}\cdot\left[{\rm Im}\left\{\mathbf{G}_{\rm B}(\mathbf{r}_{\rm a}, \mathbf{r}_{\rm a},\omega_{\rm a})\right\}\right]\cdot\mathbf{d}_{\rm a}\\
    &= \frac{\omega_{\rm a}^{3}n_{\rm B}{\rm d}^{2}}{3\pi\epsilon_{0}\hbar c^{3}}, \label{eq: GammaSEfree}
\end{align}
which accounts for non-cavity decay of the TLS. 
We highlight that all mode related coupling parameters entering the above QNM master equation are directly obtained from the QNM calculations and quantum emitter properties without any form of phenomenological fitting. We also note that, for convenience, we will refer to the quantized QNM model  as QNM-JC model, since the QNM master equation from Eq.~\eqref{eq: QNMMaster} can be viewed as a generalized and rigorous dissipative JC model.

\subsection{Diagonalization of the Lindblad dissipator}
In this subsection, we apply a unitary transformation to the QNM master equation, Eq.~\eqref{eq: QNMMaster}, to diagonalize the decay matrix $\boldsymbol{\chi}^{(-)}$ (defined below Eq.~\eqref{eq:QLE}). This will support the discussion of Section~\ref{Sec: Applications},  
and will make the role of off-diagonal QNM coupling more clear, as  it will be entirely encoded in the Hamiltonian part of the master equation. Since $\boldsymbol{\chi}^{(-)}$ is a semi-positive definite and Hermitian matrix, there exists a unitary transformation $\mathbf{U}^{(-)}$, that diagonalizes $\boldsymbol{\chi}^{(-)}$, such that 
\begin{equation}
    \sum_{\nu,\nu'}U^{(-)*}_{\nu\mu}\chi_{\nu\nu'}^{(-)}U^{(-) }_{\nu'\eta}\equiv\Gamma_\mu\delta_{\mu\eta},\label{eq: DiagChimin}
\end{equation}
where $\Gamma_\mu$ are the eigenvalues of $\boldsymbol{\chi}^{(-)}$. 
In the new basis, the Lindblad dissipator $\mathcal{L}_{\rm em}$ takes the diagonal form
\begin{equation}
    \mathcal{L}_{\rm em }\rho=\sum_{\mu}\Gamma_\mu\left(2A_\mu\rho A_\mu^\dagger - A_\mu^\dagger A_\mu\rho-\rho A_\mu^\dagger A_\mu\right),
\end{equation}
where $A_\mu^{(\dagger)}$ are QNM annihilation (creation) operator in the diagonalized dissipator frame with 
\begin{equation}
    A_\mu = \sum_{\eta}U^{(-)*}_{\eta\mu}a_\eta.\label{eq: DiagAmu}
\end{equation}

Since $U^{(-)}_{\mu\eta}$ is unitary, the bosonic commutation relations of the QMN operators are preserved. 
The effective system Hamiltonian in the diagonalized basis reads
\begin{align}
    H_{\rm sys}'=&\hbar\sum_{\mu}\Delta_\mu A_\mu^{\dagger}A_\mu + \hbar\sum_{\mu\neq\eta}g_{{\rm em}}^{\mu\eta} A_\mu^{\dagger}A_\eta  \nonumber \\
    &+ \left[\hbar\sum_\mu g_\mu A_\mu\sigma^+ + {\rm H.a.}\right]+ H_{\rm a} + H_{\rm L}, \label{eq: HamDiagPicture}
\end{align}
with $\Delta_\mu=\Omega_\mu -\omega_{\rm L}$, and  the coupling constants transform as
\begin{gather}
    \chi^{(+)}_{\mu\eta}\rightarrow \sum_{\nu,\nu'}U^{(-)*}_{\nu\mu}\chi_{\nu\nu'}^{(+)}U^{(-) }_{\nu'\eta}\equiv \tilde{\chi}^{(+)}_{\mu\eta},\\
    \tilde{g}_\mu^s\rightarrow \sum_{\nu}\tilde{g}_\nu^s U^{(-) }_{\nu\mu}\equiv g_\mu.\label{eq: gdiag}
\end{gather}

In Eq.~\eqref{eq: HamDiagPicture}, we have defined $g_{\rm em}^{\mu\eta}\equiv\tilde{\chi}^{(+)}_{\mu\eta}$ for $\mu\neq\eta$ as the photon-photon coupling constant between QNM $\mu$ and $\eta$, and $\Omega_\mu\equiv\tilde{\chi}^{(+)}_{\mu\mu}$ as the bare mode frequencies in the diagonalized and symmetrized picture. We note that $\tilde{\boldsymbol{\chi}}^{(+)}$ has the same eigenvalues as $\boldsymbol{\chi}^{(+)}$, i.e., the eigenenergies of the full photon Hamiltonian are not changed by the unitary transformation. Furthermore, the photon number operators $N_\mu = A_\mu^\dagger A_\mu$ do also not commute with the Hamiltonian in the diagonalized frame.

\subsection{Input-output relations and output electric field operator\label{Subsec: OutputEfield}}
Here, we derive the output electric field operator in the far field region and the input-output relations for multiple QNM operators (in the Heisenberg picture).
First, we write down the time-reversed quantum Langevin equation of Eq.~\eqref{eq:QLE}:
\begin{align}
\dot{a}_\mu{=}&-\frac{i}{\hbar}[a_\mu,H_{\rm sys}]+\sum_\eta\chi^{(-)}_{\mu\eta}a_\eta\nonumber\\
&-\sqrt{2}\sum_\eta\left[\left(\boldsymbol{\chi}^{(-)}\right)^{1/2}\right]_{\mu\eta}a^{\rm out}_{\eta},\label{eq:QLETR} 
\end{align}
where $a^{\rm out}_{\mu}=\int_0^\infty{\rm d}\omega a^{\rm out}_{\mu}(\omega)$ is the output operator, which is explicitly given in App.~\ref{app: InputOpApprox}, Eq.~\eqref{eq: aOutOpMarkov}. 
Next, we subtract Eq.~\eqref{eq:QLETR} from Eq.~\eqref{eq:QLE} and multiply from the left with $\left(\boldsymbol{\chi}^{(-)}\right)^{-1/2}$ to obtain the input-output relations,
\begin{equation}
    a^{\rm out}_\mu- a^{\rm in}_\mu=\sqrt{2}\sum_\eta\left[\left(\boldsymbol{\chi}^{(-)}\right)^{1/2}\right]_{\mu\eta}a_\eta.\label{eq:inout}
\end{equation}

We remark that, as a consequence of the dissipation-induced coupling, the output and input of a symmetrized QNM $\mu$ is related to a linear combination of all QNM operators, which is in contrast to the standard input-output  relation~\cite{Gardiner1}, where the input and output channel is connected via a single system operator. In the diagonalized basis, using $A_{\mu}$ instead of $a_\mu$, we obtain the (diagonalized) input-output relations:
\begin{equation}
    A^{\rm out}_\mu- A^{\rm in}_\mu=\sqrt{2\Gamma_\mu} A_\mu\label{eq:inoutDiag}.
\end{equation}
One can also formulate input-output relations in $\omega$-space of the forms 
\begin{align}
 a^{\rm out}_\mu(\omega)- a^{\rm in}_\mu(\omega)&=\sqrt{2}\sum_\eta\left[\left(\boldsymbol{\chi}^{(-)}\right)^{1/2}\right]_{\mu\eta}a_\eta(\omega)\label{eq: inout-omega_dep}, \\
    A^{\rm out}_\mu(\omega)- A^{\rm in}_\mu (\omega)&=\sqrt{2\Gamma_\mu} A_\mu(\omega),
\end{align}
where, obviously, one obtains the $\omega$-independent relations by integrating over all $\omega$ on both sides, respectively. These input-output relations will be used in the following to formulate the output electric field operator, which is important for simulations involving correlation functions at a outside detector.

We next start with the full (positive-rotating) electric field operator $\hat{\mathbf{E}}^{(+)}(\mathbf{R},t) =\int_0^\infty {\rm d}\omega\hat{\mathbf{E}}(\mathbf{R},\omega,t)$ at a position $\mathbf{R}$ outside the resonator; from the source-field expression from Eq.~\eqref{eq: SolE}, using the QNM Green function together with the field regularization (Eq.~\eqref{eq: DysonF},\eqref{eq: NF2FF_Z}), we obtain: 
\begin{align}
    \hat{\mathbf{E}}^{(+)}(\mathbf{R},t)=&i\sum_{\mu}\sqrt{\frac{\hbar\omega_{\mu} }{2\epsilon_0}}\, \int_0^\infty{\rm d}\omega\tilde{\mathbf{F}}_{\mu}^{\rm s}(\mathbf{R},\omega)a_\mu(\omega,t)\label{eq: Eoutpre1},
\end{align}
where 
\begin{equation}
    \tilde{\mathbf{F}}_{\mu}^{\rm s}(\mathbf{R},\omega)=\sum_{\eta} \tilde{\mathbf{F}}_{\eta}(\mathbf{R},\omega) \left(\mathbf{S}^{1/2}\right)_{\eta\mu}\sqrt{\frac{\omega_\eta}{\omega_\mu}} 
\end{equation}
 is a regularized QNM (Eq.~\eqref{eq: DysonF},\eqref{eq: NF2FF_Z}) in the symmetrized basis, and $a_\mu(\omega,t)$ is implicitly defined via $a_\mu(t) = \int_0^\infty {\rm d}\omega a_\mu(\omega,t)$. 
 
 Subsequently, we rewrite $\hat{\mathbf{E}}^{(+)}(\mathbf{R},t)$ from Eq.~\eqref{eq: Eoutpre1} as 
 \begin{align}
    \hat{\mathbf{E}}^{(+)}(\mathbf{R},t)=&i\sum_{\mu,\eta}\sqrt{\frac{\hbar\omega_{\mu} }{2\epsilon_0}}\, \int_0^\infty{\rm d}\omega\mathbf{F}_{\mu}^{\rm s}(\mathbf{R},\omega)\\
    &\times\sqrt{2}\left[\left(\boldsymbol{\chi}^{(-)}\right)^{1/2}\right]_{\mu\eta}a_\eta(\omega,t)\label{eq: Eoutpre2}, 
\end{align}
 with 
 \begin{equation}
     \mathbf{F}_{\mu}^{\rm s}(\mathbf{R},\omega)=\sum_\eta\mathbf{F}_{\eta}^{\rm s}(\mathbf{R},\omega)\left[\left(\boldsymbol{\chi}^{(-)}\right)^{-1/2}\right]_{\eta\mu}\sqrt{\frac{\omega_\eta}{2\omega_\mu}}.\label{eq: FregChimin}
 \end{equation}

 Now we can use the input-output relations from Eq.~\eqref{eq: inout-omega_dep}, to obtain the representation $\hat{\mathbf{E}}^{(+)}(\mathbf{R})=\hat{\mathbf{E}}^{(+)}_{\rm out}(\mathbf{R})-\hat{\mathbf{E}}^{(+)}_{\rm in}(\mathbf{R})$, where
\begin{equation}
  \hat{\mathbf{E}}_{\rm out/in}^{(+)}(\mathbf{R},t)=i\sum_\mu\sqrt{\frac{\hbar\omega_\mu}{2\epsilon_0}}\int_0^\infty {\rm d}\omega \mathbf{F}^{\rm s}_\mu(\mathbf{R},\omega)a_\mu^{\rm out/in}(\omega,t),\label{eq: OutputEOp1}
\end{equation}
are the (cavity) output and input electric field operators. 

Equation~\eqref{eq: OutputEOp1} yields a general expression for the QNM output/input field for the multi-QNM case. In the following, we concentrate on positions $\mathbf{R}$ in the far field, i.e. $|\mathbf{R}|\gg {\rm max}(\lambda_\mu)$, to obtain an approximated form of the output/input fields for numerical calculations in Section~\ref{Sec: Applications}, that connects to the $\omega$-independent QNM system operators $a_\mu$. Since $\tilde{\mathbf{F}}_\mu(\mathbf{R},\omega)$ is proportional to the background Green function $\mathbf{G}_{\rm B}(\mathbf{R},\mathbf{r},\omega)$, where $\mathbf{r}$ is either located in the resonator volume (Eq.~\eqref{eq: DysonF}) or at the resonator boundary (Eq.~\eqref{eq: NF2FF_Z}), i.e., $|\mathbf{R}|\gg |\mathbf{r}|$, we can approximate the regularized function as
\begin{equation}
    \tilde{\mathbf{F}}_\mu(\mathbf{R},\omega)\approx \tilde{\mathbf{Z}}_\mu(\mathbf{R},\omega) e^{in_{\rm B}\omega|\mathbf{R}|/c},\label{eq: ApproxFreg}
\end{equation}
where $\tilde{\mathbf{Z}}_\mu(\mathbf{R},\omega)$ for
the Ansatz in Eq.~\eqref{eq: NF2FF_Z}, is given as
\begin{align}
\begin{split}
&\tilde{\mathbf{Z}}_\mu(\mathbf{R},\omega)=i\omega\mu_0\frac{1}{4\pi|\mathbf{R}|}\oint_{\mathcal{S}'} {\rm d}S' e^{-in_{\rm B}\omega\hat{\mathbf{R}}\cdot\mathbf{s}'/c}\\
& \ \ \ \ \times\bigg[\tilde{\mathbf{J}}_\mu(\mathbf{s}') - \left(\tilde{\mathbf{J}}_\mu(\mathbf{s}')\cdot\hat{\mathbf{R}}\right)\hat{\mathbf{R}}
-n_{\rm B}c\epsilon_0\hat{\mathbf{R}}\times\tilde{\mathbf{M}}_\mu(\mathbf{s}')\bigg],\label{eq: NF2FF_ZApprox}
\end{split}
\end{align}
and $\hat{\mathbf{R}}=\mathbf{R}/|\mathbf{R}|$ is the unit vector in the direction of $\mathbf{R}$. 

Inserting Eq.~\eqref{eq: ApproxFreg} into the output/input electric field operator from Eq.~\eqref{eq: OutputEOp1},  yields 
\begin{align}
    \hat{\mathbf{E}}_{\rm out/in}^{(+)}(\mathbf{R},t)\approx & i\sum_\mu\sqrt{\frac{\hbar\omega_\mu}{2\epsilon_0}}\int_0^\infty {\rm d}\omega \mathbf{Z}^{\rm s}_\mu(\mathbf{R},\omega)\nonumber\\
    &\times e^{in_{\rm B}\omega|\mathbf{R}|/c}a_\mu^{\rm out/in}(\omega,t),\label{eq: OutputEOp2}
\end{align}
where $\mathbf{Z}^{\rm s}_\mu(\mathbf{R},\omega)$ is implicitly defined via Eq.~\eqref{eq: FregChimin} together with Eq.~\eqref{eq: ApproxFreg} and \eqref{eq: NF2FF_ZApprox}.
Using the definition of $a_\mu^{\rm out/in}(\omega,t)$ (Eq.~\eqref{eq: aInOpMarkov} and Eq.~\eqref{eq: aOutOpMarkov}), it follows that $ e^{in_{\rm B}\omega|\mathbf{R}|/c}a_\mu^{\rm out/in}(\omega,t)=a_\mu^{\rm out/in}(\omega,t-n_{\rm B }|\mathbf{R}|/c)$. Since $\mathbf{Z}^{\rm s}_\mu(\mathbf{R},\omega)$ varies slowly with respect to $\omega$ around the QNM frequency $\omega_\mu$, we apply a resonance approximation to obtain the final expression
for the output field operator:
\begin{align}
    \hat{\mathbf{E}}_{\rm out/in}^{(+)}(\mathbf{R},t)\approx  i\sum_\mu\sqrt{\frac{\hbar\omega_\mu}{2\epsilon_0}}\mathbf{Z}^{\rm s}_\mu(\mathbf{R}) a_\mu^{\rm out/in}(t-n_{\rm B}|\mathbf{R}|/c),\label{eq: Eout2}
\end{align}
where we used again $a_\mu^{\rm out/in}=\int_0^\infty {\rm d}\omega a_\mu^{\rm out}(\omega)$. In the diagonalized basis, the output/input field reads 
\begin{align}
    \hat{\mathbf{E}}_{\rm out/in}^{(+)}(\mathbf{R},t)=  i\sum_\mu\sqrt{\frac{\hbar\omega_\mu}{2\epsilon_0}}\mathbf{Z}^{\rm sU}_\mu(\mathbf{R}) A_\mu^{\rm out}(t-n_{\rm B}|\mathbf{R}|/c),\label{eq: Eout3}
\end{align}
with
\begin{equation}
    \mathbf{Z}_{\mu}^{\rm sU}(\mathbf{R})=\sum_{\eta,\eta'} \tilde{\mathbf{Z}}_{\eta}(\mathbf{R},\omega_\eta) \left(\mathbf{S}^{1/2}\right)_{\eta\eta'}U_{\mu\eta'}^{(-)}\sqrt{\frac{\omega_\eta}{2\omega_\mu\Gamma_\mu}}\label{eq: FregDiagU}.
\end{equation}

Using the $\omega$-independent input-output relations from Eq.~\eqref{eq:inout} or Eq.~\eqref{eq:inoutDiag}, we can connect the output field to the system QNM operators $a_\mu$ or $A_\mu$. In particular, the output electric field operator at position $\mathbf{R}$ and time $t$ is then a linear combination of far-field regularized QNMs $\mathbf{Z}_{\mu}^{\rm s}(\mathbf{R}) (\mathbf{Z}_{\mu}^{\rm sU}(\mathbf{R}))$ and QNM system operators $a_\mu (A_\mu)$ as well as QNM input operators $a^{\rm in}_\mu (A^{\rm in}_\mu)$ at time $t-n_{\rm B}|\mathbf{R}|/c$.
The introduction of the above output electric field operators (Eq.~\eqref{eq: Eout2} and \eqref{eq: Eout3}) allows one to calculate, e.g., second-order photon correlation functions.

\section{Applications to coupled open resonators\label{Sec: Applications}}
In this section, we will apply the theory from Section~\ref{Sec: Theor} to a two-QNM multiphoton system, using first principle calculations for a specific 
open cavity structure. A typical system to study in terms of two dominant but different QNMs are metal-dielectric hybrid structures, where one mode is photon dominated and one is plasmon dominated~\cite{KamandarDezfouli2017,koenderink2010use,palstra2019hybrid}, but with a sufficiently different quality factor.

We will focus on  the hybrid metal-dielectric structure depicted in Fig.~\ref{fig: Scheme}, which shows two fundamental QNMs in the optical frequency regime. In particular, we will discuss differences between a phenomenological dissipative JC model and the QNM-JC 
model with respect to density matrix equations results of the hybrid structure. In subsection~\ref{subseq: 2ModeMaster}, we will discuss the Hamiltonian and the dissipator of the master equations on a formal basis. Afterwards, both models (phenomenological dissipative JC and QNM-JC) will be compared in the weak photon-emitter coupling regime in subsection~\ref{subsec: PF}, and subsequently in the strong emitter-photon coupling regime in subsections~\ref{subsec: sysprop} and \ref{subsec: Outputprop}.

\subsection{Two-QNM master equations, hybrid cavity  and TLS parameters\label{subseq: 2ModeMaster}}
We start with the formulation of the master equation for the hybrid structure, e.g., as shown in Fig.~\ref{fig: Scheme}. All input QNM and TLS parameters used for numerical evaluation can be found on the fourth row of Tab.~\ref{Tab: Params}. A more detailed description of the QNM input parameters is given in App.~\ref{app: OG_Params}.

For the two-mode hybrid case, we rewrite the Hamiltonian  
for the QNM-JC model $H_{\rm sys}^{\prime\rm QNM}\equiv H_{\rm sys}'$ from Eq.~\eqref{eq: HamDiagPicture} with $\mu=\{{\rm pc,pl}\}$ as 
\begin{align}
   H_{\rm sys}^{\prime\rm QNM}=& \hbar\Delta_{\rm pl}A^{\dagger}_{\rm pl}A_{\rm pl} + \hbar\Delta_{\rm pc}A^{\dagger}_{\rm pc}A_{\rm pc} + \hbar\Delta_{\rm a}\sigma^+\sigma^- \nonumber\\
   &+ \hbar \left[g_{\rm pl}\sigma^+ A_{\rm pl} +g_{\rm pc}\sigma^+ A_{\rm pc} + {\rm H.a.}\right]+ H_{\rm L}'\nonumber\\
   &+\hbar\left[g_{\rm em}A^{\dagger}_{\rm pl} A_{\rm pc} +{\rm H.a.}\right],\label{eq: HsysTwoModeQNMJC}
\end{align}
where $A^{(\dagger)}_{\rm pl}$ is the annihilation (creation) operator for the plasmon-like mode and $A^{(\dagger)}_{\rm pc}$ is the annihilation (creation) operator for the PC-like mode in the diagonalized dissipator picture, as illustrated in Fig.~\ref{fig: Scheme} (c-d). 

In this basis, $\Delta_{\rm pl(pc)}=\Omega_{\rm pl(pc)}-\omega_{\rm L}$ is the detuned plasmon-like (PC-like) frequency; $g_{\rm pl}$ and $g_{\rm pc}$ are the coupling constants of the plasmon and PC mode to the TLS, respectively. To simplify the notation, we have set $g_{\rm em}^{\rm pl,pc}=g_{\rm em}$ as the plasmon-PC mode coupling constant.
The Lindblad dissipator is then
\begin{equation}
    \mathcal{L}_{\rm diss}^{\rm QNM}\rho = \Gamma_{\rm pl}\mathcal{D}[A_{\rm pl}]\rho + \Gamma_{\rm pc}\mathcal{D}[A_{\rm pc}]\rho+ \mathcal{L}_{\rm SE}\rho,
\end{equation}
with 
\begin{equation}
    \mathcal{D}[A]\rho = 2A\rho A^\dagger - A^\dagger A \rho - \rho A^\dagger A ,\label{eq: GenLindbladterm}
\end{equation}
and where $\Gamma_{\rm pl(pc)}$ is the diagonalized decay rate of the PC-like (plasmon-like) QNM, defined implicitly from Eq.~\eqref{eq: DiagChimin}.

\begin{table}[htb]
\caption {Computed mode frequencies $\Omega_{\rm pl/pc}$ ($\omega_{\rm pl/pc}$), decay rates $\Gamma_{\rm pl/pc}$ ($\gamma_{\rm pl/pc}$), photon-emitter coupling constants $g_{\rm pl/pc}$ ($\tilde{g}_{\rm pl/pc}$), as well as photon-photon coupling constants $g_{\rm em}$ ($\tilde{g}_{\rm em}$) for the hybrid in Fig.~\ref{fig: Scheme} using the QNM-JC (phenomenological dissipative JC) model.} \label{Tab: Params} 
    \centering
    \begin{tabular}{|c|c|c|c|}
 \hline
   & QNM-JC model & & JC model \\
 \hline
 $\hbar\Omega_{\rm pl}~[{\rm eV}]$ & $1.6988$ & $\hbar\omega_{\rm pl}~[{\rm eV}]$ &  $1.6999$ \\
 \hline
 $\hbar\Omega_{\rm pc}~[{\rm eV}]$ & $1.6063$ & $\hbar\omega_{\rm pc}~[{\rm eV}]$ & $1.6052$\\
 \hline
 $\hbar\Gamma_{\rm pl}~[{\rm meV}]$ & $48.5$ & $\hbar\gamma_{\rm pl}~[{\rm meV}]$ & $47.9$ \\
 \hline
 $\hbar\Gamma_{\rm pc}~[{\rm meV}]$ & $0.1$ & $\hbar\gamma_{\rm pc}~[{\rm meV}]$ & $0.7$ \\
 \hline
 $i\hbar g_{\rm pl}~[{\rm meV}]$ & $45.7-0.9i$ & $i\hbar\tilde{g}_{\rm pl}~[{\rm meV}]$ & $46.5-1.2i$\\
 \hline
 $i\hbar g_{\rm pc}~[{\rm meV}]$ & $0.7 + 0.6i$ & $i\hbar\tilde{g}_{\rm pc}~[{\rm meV}]$ & $5.3+2.4i$\\
 \hline
 $\hbar g_{\rm em}~[{\rm meV}]$ & $-11.6+7.3i$ & $\hbar\tilde{g}_{\rm em}$ & $0$
  \\
  \hline
    \end{tabular}
\end{table}

In contrast, a phenomenological dissipative JC model assuming $[\tilde{\alpha}_\mu,\tilde{\alpha}_\eta^\dagger]=\delta_{\mu\eta}$ for $\mu,\eta=1,2$ is represented by the Hamiltonian 
\begin{align}
   \tilde{H}^{\prime\rm JC}_{\rm sys}=& \hbar\tilde{\Delta}_{\rm pl}\tilde{\alpha}^{\dagger}_{\rm pl}\tilde{\alpha}_{\rm pl} + \hbar\tilde{\Delta}_{\rm pc}\tilde{\alpha}^{\dagger}_{\rm pc}\tilde{\alpha}_{\rm pc} + \hbar\Delta_{\rm a}\sigma^+\sigma^- \nonumber\\
   &+ \hbar \left[\tilde{g}_{\rm pl}\sigma^+ \tilde{\alpha}_{\rm pl} +\tilde{g}_{\rm pc}\sigma^+ \tilde{\alpha}_{\rm pc}  + {\rm H.a.}\right]+ H_{\rm L}',\label{eq: HsysTwoModeJC}
\end{align}
with $\tilde{g}_{\rm pl(pc)}=-i\sqrt{\omega_{\rm pl(pc)}/(2\hbar\epsilon_0)}\mathbf{d}_{\rm a}\cdot\tilde{\mathbf{f}}_{\rm pl(pc)}(\mathbf{r}_{\rm a})$ using the untransformed QNM fields of the hybrid and $ \tilde{\Delta}_{\rm pl(pc)} = \omega_{\rm pl(pc)} - \omega_{\rm L}$. The Lindblad dissipator reads
\begin{equation}
    \tilde{\mathcal{L}}_{\rm diss}^{\rm JC}\rho = \gamma_{\rm pl}\mathcal{D}[\tilde{\alpha}_{\rm pl}]\rho  + \gamma_{\rm pc}\mathcal{D}[\tilde{\alpha}_{\rm pc}]\rho +\mathcal{L}_{\rm SE}\rho,
\end{equation}
and we note that $\omega_\mu -i\gamma_\mu$ ($\mu=\{{\rm pl,pc}\}$) are the real and imaginary part of the original, individual hybrid QNM eigenfrequencies, respectively. We emphasize that the Lindblad dissipator $\mathcal{L}_{\rm SE}\rho$ associated to the vacuum spontaneous emission rate $\gamma_{\rm SE}$ of the TLS is the same in both models.

The corresponding master equations are given by 
\begin{equation}
   \partial_t\rho = -\frac{i}{\hbar}[H^{\prime\rm QNM}_{\rm sys},\rho]+\mathcal{L}_{\rm diss}^{\rm QNM}\rho,\label{eq: MasterTwoModeQNMJC}
\end{equation}
or
\begin{equation}
    \partial_t\rho = -\frac{i}{\hbar}[\tilde{H}^{\prime\rm JC}_{\rm sys},\rho]+\tilde{\mathcal{L}}_{\rm diss}^{\rm JC}\rho,\label{eq: MasterTwoModeJC}
\end{equation}
for the QNM-JC model and phenomenological dissipative JC model, respectively. 

Next, to specify the differences between both master equations, namely Eqs.~(\ref{eq: MasterTwoModeQNMJC},\ref{eq: MasterTwoModeJC}), we compare the occurring coupling parameters in the two different master equations for the same hybrid structure, which are summarized in Tab.~\ref{Tab: Params}. 
We see that the overall behaviour of the dominating plasmon-mode related quantities are very similar before (phenomenological dissipative JC model) and after symmetrization and diagonalization (QNM-JC model). However, some of the PC-mode related parameters change drastically, as described below:

(i) \textit{Dissipation.}---The effective width $\Gamma_{\rm pc}$ of the symmetrized PC-QNM is around one order of magnitude smaller compared to the original PC width $\gamma_{\rm pc}$. This is a consequence of the structure of the decay matrix $\boldsymbol{\chi}^{(-)}$(cf.~the text surrounding Eq.~\eqref{eq:QLE}), since its elements are linear combination of the original complex eigenfrequencies $\tilde{\omega}_\mu$.
Indeed, for two modes, the exact decay eigenvalues $\Gamma_{\rm pl/pc}$ for the plasmon mode and the PC mode are
\begin{align}
    \Gamma_{\rm pl/pc}=\frac{{\rm tr}\left(\boldsymbol{\chi}^{(-)}\right)}{2}\pm \sqrt{\frac{\left[{\rm tr}\left(\boldsymbol{\chi}^{(-)}\right)\right]^2}{4}-{\rm det}\left(\boldsymbol{\chi}^{(-)}\right)}\, ,
\end{align}   
and, using the properties of the trace and the determinant,
\begin{equation}
    \Gamma_{\rm pl/pc}=\frac{\gamma_{\rm pl}\left[1\pm \sqrt{1+R}\right]}{2}+\frac{\gamma_{\rm pc}\left[1\mp \sqrt{1+R}\right]}{2}\label{eq: EigVal}.
\end{equation}

Here, $R$ is defined through    
\begin{equation}
    R = \frac{ |S_{\rm pl,pc}|^2|(\tilde{\omega}_{\rm pl}-\tilde{\omega}_{\rm pc}^*)|^2 -4S_{\rm pl,pl}S_{\rm pc,pc}\gamma_{\rm pl}\gamma_{\rm pc} }{{\rm det}\left[\mathbf{S}\right](\gamma_{\rm pl}-\gamma_{\rm pc})^2},
\end{equation}
which is  a dissipation-induced correction factor to the initial phenomenological damping. 

The analytic form of the eigenvalues in Eq.~\eqref{eq: EigVal} has an interesting implication: if $\gamma_{\rm pl}$ and $\gamma_{\rm pc}$ are very different from each other, e.g., $\gamma_{\rm pl}\gg\gamma_{\rm pc}$, 
then the plasmon-related eigenvalue $\Gamma_{\rm pl}$ is only slightly shifted compared to $\gamma_{\rm pl}$, since the correction by the term involving $\gamma_{\rm pc}$ is very small. In contrast, $\Gamma_{\rm pc}$ is mainly influenced by the correction term corresponding to $\gamma_{\rm pl}$. On the other hand, when $\gamma_{\rm pc}\sim \gamma_{\rm pl}\equiv\gamma$, then we find, as a first estimate (for a very small difference below 10\%),  
\begin{equation}
    \Gamma_{\rm pl/pc}\approx\gamma \pm |S_{\rm pl,pc}|\frac{\omega_{\rm pl}-\omega_{\rm pc}}{2\sqrt{{\rm det}\left(\mathbf{S}\right)}}.\label{eq: DissEigSpecialCase}
\end{equation}
Thus there is a symmetric splitting of $\gamma$, which depends on the detuning of both modes and the mode overlap $|S_{\rm pl,pc}|$. Strictly speaking, Eq.~\eqref{eq: DissEigSpecialCase} is exact for degenerate QNM imaginary parts $\gamma_{\rm pl}=\gamma_{\rm pc}$, which is a technical interesting case, since it appears, e.g., in Fabry-Pérot cavities.

In the example of Fig.~\ref{fig: Scheme}, the former case $\gamma_{\rm pl}\gg \gamma_{\rm pc}$ applies, which explains why the effective PC-mode decay rate $\Gamma_{\rm pc}$ is significantly shifted 
from $\gamma_{\rm pc}$ (cf. Tab.~\ref{Tab: Params}). It should be noted that these rate changes  are also present in the full eigenvalues of the Liouvillians, which will be important for the response of the hybrid to an external optical field, as we will discuss later. 

(ii) \textit{QNM-TLS coupling.}---Another interesting observation is that the PC-TLS coupling constant $g_{\rm pc}$ in the QNM JC-model (Eq.~\eqref{eq: HsysTwoModeQNMJC}) is also nearly one order magnitude lower compared to the phenomenological dissipative JC parameter $\tilde{g}_{\rm pc}$ (Eq.~\eqref{eq: HsysTwoModeJC}), using the original PC-mode eigenfunction $\tilde{\mathbf{f}}_\mu$; this is because $g_{\rm pc}$ is formed by a linear combination of $\tilde{g}_{\rm pc}$ and $\tilde{g}_{\rm pl}$, which also deviate by one order of magnitude to each other.
Interestingly, since the effective PC decay rate changes by a similar amount, both parameter sets lead to nearly the same photon-emitter coupling to decay ratio, i.e., $|g_{\rm pc}|/(2\Gamma_{\rm pc})\approx |\tilde{g}_{\rm pc}|/(2\gamma_{\rm pc})\approx 4$ in the QNM-JC model as well as in the phenomenological dissipative JC model (cf. Eq.~\eqref{eq: HsysTwoModeQNMJC} and \eqref{eq: HsysTwoModeJC}).

To summarize this subsection, there are two key changes that occur by comparing the two master equations, Eq.~\eqref{eq: MasterTwoModeQNMJC},\eqref{eq: MasterTwoModeJC}.  We first recognize {\it modifications of the mode and coupling parameters (mainly PC mode) due to symmetrization and diagonalization of the QNM annihilation and creation operators}, and second, there are {\it additional contributions in the Hamiltonian due to the presence of the photon-photon interaction part with the coupling constant $g_{\rm em}$} (Eq.~\eqref{eq: HsysTwoModeQNMJC}).

The impact of these key changes, 
which results from a proper treatment of the QNM quantization (with real losses), will be shown below by 
explicitly comparing the phenomenological dissipative JC and QNM master equation simulations. First, in subsection~\ref{subsec: PF}, we inspect the weak light-electron coupling regime, where we apply the bad cavity limit as in Ref.~\onlinecite{PhysRevLett.122.213901} (but for a completely different hybrid structure), and adiabatically eliminate the cavity modes from the master equations, Eqs.~\eqref{eq: MasterTwoModeQNMJC},\eqref{eq: MasterTwoModeJC}. In this limit, we compare the cavity-enhanced spontaneous emission of the quantum emitter obtained from the QNM-JC model, the phenomenological dissipative JC-model and an independent semi-classical solution. 
Subsequently, in subsection~\ref{subsec: sysprop} and \ref{subsec: Outputprop}, we analyse the multiphoton regime in the strong light-electron coupling regime, where the bad cavity approximations are not valid anymore.  
Thus, we will use the full master equations, Eqs.~\eqref{eq: MasterTwoModeQNMJC},\eqref{eq: MasterTwoModeJC}. The numerical results of the master equations were calculated using the library Quantum Toolbox in Python~\cite{johansson2013qutip} (QuTiP) and we note again, that all parameters, which enter the master equations are summarized in Tab.~\ref{Tab: Params}.

\subsection{Weak light-exciton coupling regime: Purcell factors and radiative $\beta$ factors\label{subsec: PF}}
\begin{figure}[!h]
	\centering
	\includegraphics[width=0.97\columnwidth,trim={0.20cm 0 0.15cm 0},clip]{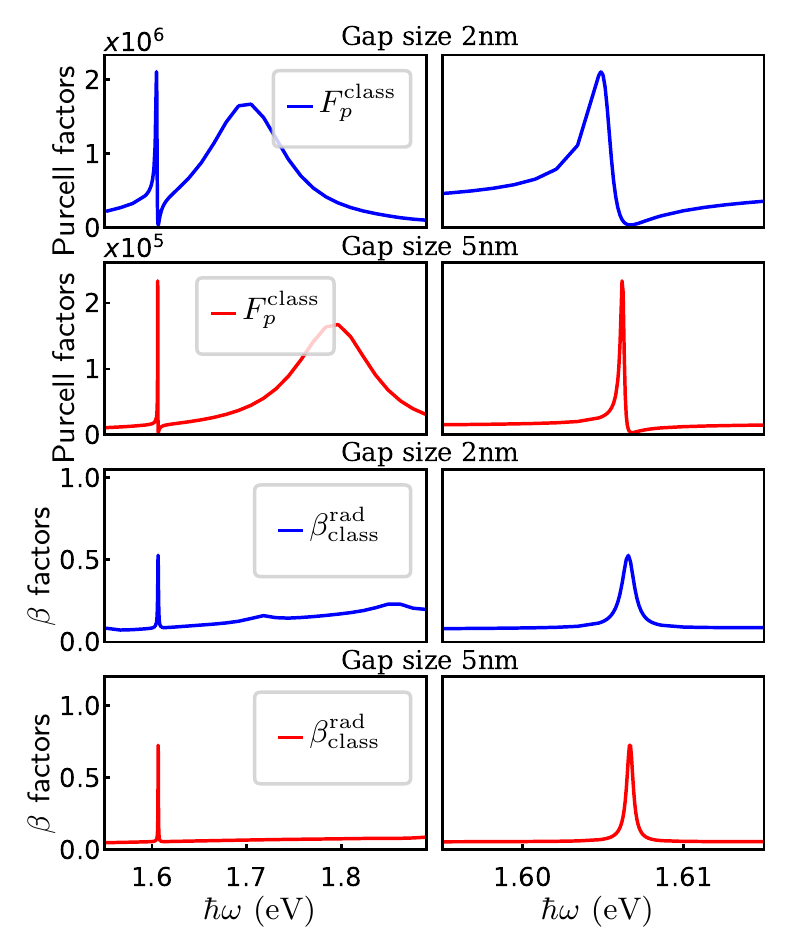}
	\caption{Classical Purcell factors (first and second row) and classical radiative $\beta$-factors (third and fourth row)
	from an embedded dipole emitter in the metal-dielectric hybrid structure from Fig.~\ref{fig: Scheme} with gap size of 2 nm and 5 mn, obtained from full Maxwell simulations over a frequency regime covering both hybrid resonances (left) with a zoom-in close to the high-$Q$ resonance (right). Note that for the 5 nm (2nm) gap case, the numerical calculations were done for 81 (93) non-equidistant frequency points in the interval $\hbar\omega\in [1.55,1.95]~$eV, and that the corresponding $F_p^{\rm num}$ and $\beta^{\rm rad}_{\rm class}$ values are linearly interpolated. Also note  that the maximum Purcell factor of the 2 nm gap hybrid is one order of magnitude higher compared to the 5 nm gap hybrid. 
	The value of 
	$\beta^{\rm rad}_{\rm class}$ for the 2 nm gap has additional peaks for frequencies towards the low-$Q$ mode, which likely come from non-modal quasi-static coupling and a constant background term with higher order modes, whose influence reduces with decreasing gap size.
	For all simulations, the classical dipole is at the
	gap center of the dimer and is $z$-polarized. 
	}\label{fig: beta}
\end{figure}

In a first step, we compare the QNM-JC model and the phenomenological dissipative JC model in the weak coupling limit, 
i.e., we reduce the QNM-emitter constants by choosing a small dipole moment ($d_\mathrm{a}<0.1~D$) of the TLS and leaving the mode parameters unchanged (cf. App.~\ref{app: CouplingRegime} on the discussion of the coupling regimes). 
In the weak coupling limit, we adiabatically eliminate both modes from the master equations, Eq.~(\ref{eq: MasterTwoModeQNMJC}-\ref{eq: MasterTwoModeJC}) (as in Ref.~\onlinecite{PhysRevLett.122.213901}) to obtain the master equations in the bad cavity limit~\cite{Cirac},
\begin{align}
    \partial_t\rho_{\rm a}^{\rm QNM} =& -\frac{i}{\hbar}\left[H_{\rm a}'+H_{\rm L}',\rho^{\rm QNM}_{\rm a}\right]\\
    &+\frac{\gamma_{\rm SE}}{2}\mathcal{D}[\sigma^-]\rho_{\rm a}^{\rm QNM}+\frac{\Gamma}{2}\mathcal{D}[\sigma^-]\rho_{\rm a}^{\rm QNM},\label{eq: BadMEQNM}
\end{align}
and 
\begin{align}
    \partial_t\tilde{\rho}_{\rm a}^{\rm JC} =& -\frac{i}{\hbar}\left[H_{\rm a}'+H_{\rm L}',\tilde{\rho}^{\rm JC}_{\rm a}\right]\\
    &+\frac{\gamma_{\rm SE}}{2}\mathcal{D}[\sigma^-]\tilde{\rho}_{\rm a}^{\rm JC}+\frac{\tilde{\Gamma}}{2}\mathcal{D}[\sigma^-]\tilde{\rho}_{\rm a}^{\rm JC},\label{eq: BadMEJC}
\end{align}
for the QNM-JC model and phenomenological dissipative JC model, respectively. Here, $\Gamma$  and $\tilde{\Gamma}$
are the cavity enhanced spontaneous emission rates of the TLS, defined through~\cite{PhysRevLett.122.213901} 
\begin{gather}
    \Gamma=\sum_{\mu,\eta={\rm pc,pl}}S_{\mu\eta}\tilde{g}_\mu \tilde{g}_\eta^*\frac{i(\omega_\mu-\omega_\eta)+\gamma_\mu+\gamma_\eta}{(\Delta_{\mu a}-i\gamma_\mu)(\Delta_{\eta a}+i\gamma_\eta)},\label{eq: GammaEnhQNM}
    \end{gather}
    and
    \begin{gather}
    \tilde{\Gamma}=\sum_{\mu={\rm pc,pl}}\frac{2|\tilde{g}_\mu|^2\gamma_\mu}{\Delta_{\mu a}^2+\gamma_\mu^2}\label{eq: GammaEnhJC},
\end{gather}
where $\Delta_{\mu a}=\omega_\mu-\omega_{\rm a}$ is the QNM-TLS detuning. 

We stress again that both models use the same original QNM parameter ($\omega_\mu,\gamma_\mu,\tilde{g}_\mu$). From the above equations, we clearly see, that in the limit $\mathbf{S}\rightarrow \mathbf{1}$, i.e., when there is vanishing radiative and non-radiative QNM overlap (cf. Eq.~\eqref{eq: Scommute}), $\tilde{\Gamma}$ and $\Gamma$ coincide. However, we should note here, that this is really only the case, when $\gamma_\mu\rightarrow 0$, which in a sense contradicts with the assumptions of a phenomenological dissipative JC model for finite loss.

Comparing the QNM-JC model and the phenomenological dissipative JC model with respect to the bad cavity limit master equations (Eqs.~\eqref{eq: BadMEQNM},\eqref{eq: BadMEJC}), we see that the differences can be summarized as additional off-diagonal terms in the QNM-JC cavity-enhanced spontaneous emission rate $\Gamma$ of the TLS, Eq.~\eqref{eq: GammaEnhQNM}. In contrast, in the full master equations (Eqs.~\eqref{eq: MasterTwoModeQNMJC},\eqref{eq: MasterTwoModeJC}), the differences of both models is not only present in the coupling constants, but also off-diagonal coupling between the different mode operator appear in the QNM-JC model.

For a demonstration of the influence of the QNM coupling terms in Eq.~\eqref{eq: GammaEnhQNM}, we calculate the Purcell factor $F_{P}=\Gamma/\gamma_{\rm SE}$  and $\tilde{F}_{P}=\tilde{\Gamma}/\gamma_{\rm SE}$ (cf.~Eq.~\eqref{eq: GammaSEfree}) as a function of the TLS frequency assuming the approximative bad cavity limit~\cite{PhysRevLett.122.213901},  for the QNM master equation and the two-mode phenomenological dissipative JC master equation. To estimate the quality of the Purcell factor results and the underlying models in the bad cavity limit, we compare these results with a (independent) semi-classical Maxwell simulation. 

However,  before comparing the two different quantum models and semi-classical model for the specific hybrid in Fig.~\ref{fig: Scheme}, we first discuss the choice for the design of the hybrid structure using results obtained solely from the full Maxwell simulations. To do so, we compare the hybrid design from Fig.~\ref{fig: Scheme} with a gap of ellipsoidal dimer of 2 nm and 5 nm with respect to the Purcell factor and radiative $\beta$-factor in Fig.~\ref{fig: beta}. 

The classical Purcell factor $F_{p}^{\rm class}$ is defined via
\begin{equation}
    F_{p}^{\rm class}=\frac{\oint_{ \mathcal{S}_{\rm dipole}}\hat{\mathbf{n}}\cdot {\bf S}_{\rm dipole,total}(\mathbf{r},\omega)d{\rm A} }{\oint_{\mathcal{S}_{\rm  dipole}}\hat{\mathbf{n}}\cdot {\bf S}_{\rm dipole,background}(\mathbf{r},\omega)d{\rm A} },\label{eq: PFnum}
\end{equation}
where $\mathcal{S}_{\rm dipole}$ is a small spherical surface (with radius smaller then half of the gap width) surrounding the dipole point and $\hat{\mathbf{n}}$ is a unit vector normal to $\mathcal{S}_{\rm dipole}$, pointing outward.
The vector ${\bf S}(\mathbf{r},\omega)$ is the Poynting vector at $\mathcal{S}_{\rm dipole}$ and the subscripts `total' and `background' represent the case with and without resonator.
Furthermore, the classical radiative beta factor $\beta^{\rm rad}_{\rm class}$ is defined as 
\begin{gather}
    \beta_{\rm class}^{\rm rad}=\frac{\oint_{\mathcal{S}_{\rm PML}}\hat{\mathbf{n}}\cdot {\bf S}_{\rm PML,total}(\mathbf{r}_{\rm PML},\omega)d{\rm A} }{\oint_{\mathcal{S}_{\rm dipole}}\hat{\mathbf{n}}\cdot {\bf S}_{\rm dipole,total}(\mathbf{r},\omega)d{\rm A} },\label{eq: betaradnum}
\end{gather}
where the surface $\mathcal{S}_{\rm PML}$ is the interface 
just before the PML (perfectly matched layers), surrounding the resonator structure,
and the vector ${\bf S}_{\rm PML,total}$ is the Poynting vector at $\mathcal{S}_{\rm PML}$. We emphasize, that for decreasing gap size, $\beta_{\rm class}^{\rm rad}$ is very sensitive to the choice of the two surfaces $\mathcal{S}_{\rm PML}$ and $\mathcal{S}_{\rm dipole}$, which can lead to an increased numerical uncertainty of the $\beta_{\rm class}^{\rm rad}$ calculations.  

For the above defined quantities, there are two main differences between the 2 nm and 5 nm gap size case: First, as shown in Figure~\ref{fig: beta} (first and second row), the maximum Purcell factor of the hybrid structure with 2 nm gap is roughly one order of magnitude higher compared to the 5 nm gap hybrid. Second, the $\beta$-factor (cf. Fig.~\ref{fig: beta} (third and fourth row)) shows a constantly increasing contribution with additional resonances for frequencies towards the low-$Q$ hybrid resonance for the 2 nm gap case, which are caused by effects beyond the two hybrid resonance description. Therefore, we have chosen an extreme case for the hybrid design for the quantum simulations, in that the gap between the two metallic ellipsoids is only 2 nm (cf. Fig~\ref{fig: Scheme}). In such a small gap regime, one expects additional effects beyond the main two modes, and also the numerical calculations of the total beta factor are more difficult, as mentioned above.
\begin{figure}[!h]
	\centering
	\includegraphics[trim={0.20cm 0 0.15cm 0},clip]{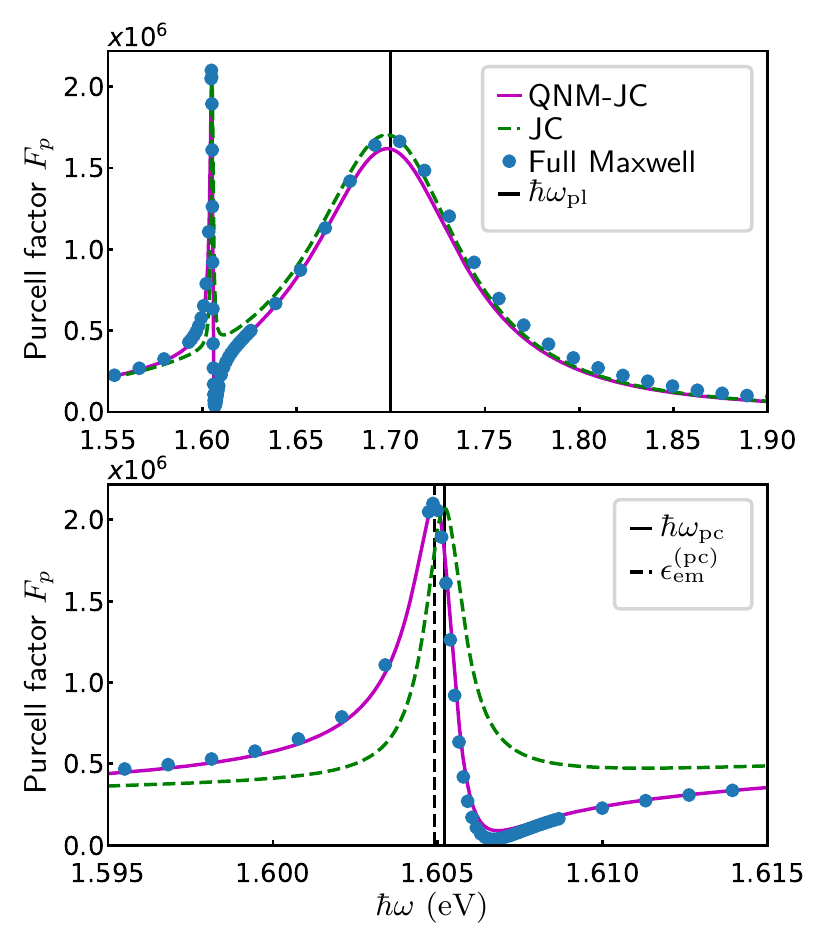}
	\caption{
	Purcell factor of a $z$-polarized dipole emitter in the metal-dielectric hybrid structure from Fig.~\ref{fig: Scheme}, shown over  a broad frequency range from full dipole simulations (dots), the two-mode QNM-JC  model (solid magenta line) and the phenomenological dissipative JC model (dashed green line). Two peaks appear around the hybridized QNM frequencies $\hbar\tilde{\omega}_{\rm pl}=1.6999 - 0.0479i~({\rm eV)}$ and $\hbar\tilde{\omega}_{\rm pc}=1.6052 - 0.0007i~({\rm eV)}$, originating from the metallic ellipsoidal dimer and the photonic-crystal beam, respectively. The highly non-Lorentz interference effect (dip in $F_p$) is located near one of the eigenfrequencies, $\epsilon_{\rm em}^{(\rm pc)}$, of the full electromagnetic Hamiltonian $H_{\rm em}$ of the QNM-JC model and is fully recaptured by the inter-mode coupling terms. }\label{fig: PF_quant_Hybrid}
\end{figure}
We choose such an extreme case, since it is more interesting for cavity-QED and emerging experiments~\cite{Barth2010,chikkaraddy2016single}, since it allows one to reach the strong coupling regime for realistic dipole strengths of the TLS, because of the large Purcell factor.

After having discussed the choice for the hybrid design in Fig.~\ref{fig: Scheme}, we next compare the Purcell factors for the QNM-JC model and phenomenological dissipative JC model with respect to the semi-classical Maxwell solution.
As shown in Fig.~\ref{fig: PF_quant_Hybrid} the QNM master equation result is in very good agreement with the full but semi-classical Maxwell simulations and reproduces the pronounced Fano effect near the original PC-like mode frequency $\omega_{{\rm pc}}$. Notably, the peak of the Purcell enhancement of the PC-like cavity mode is located at the eigenvalue $\epsilon_{\rm em}^{(\rm pc)}$ of the electromagnetic part of the Hamiltonian $H_{\rm em}$, in agreement with the derivation for the hybrid structure in Ref.~\onlinecite{PhysRevLett.122.213901}. While the Purcell factor obtained from the phenomenological dissipative JC-model is in good agreement with the full Maxwell solution near the plasmonic-like mode frequency $\omega_{\rm pl}$, it fails in the frequency regime where interference occurs. This is caused by the missing inter-mode coupling terms in the phenomenological dissipative JC model~\cite{PhysRevLett.122.213901} (Eq.~\eqref{eq: GammaEnhJC}), which is present in the QNM-JC model. It should be noted, that $S_{22}$ was decreased from $1$ to $0.77$ in the phenomenological model to at least match the height of the phenomenological dissipative JC result at the PC peak. This leads to a slightly modified PC-TLS coupling constant for the phenomenological dissipative JC model of $\tilde{g}_{2}\rightarrow \sqrt{0.77}\tilde{g}_{2}$, but all other parameters are identical.
\begin{figure}[!h]
	\centering
	\includegraphics[trim={0.20cm 0 0.15cm 0},clip]{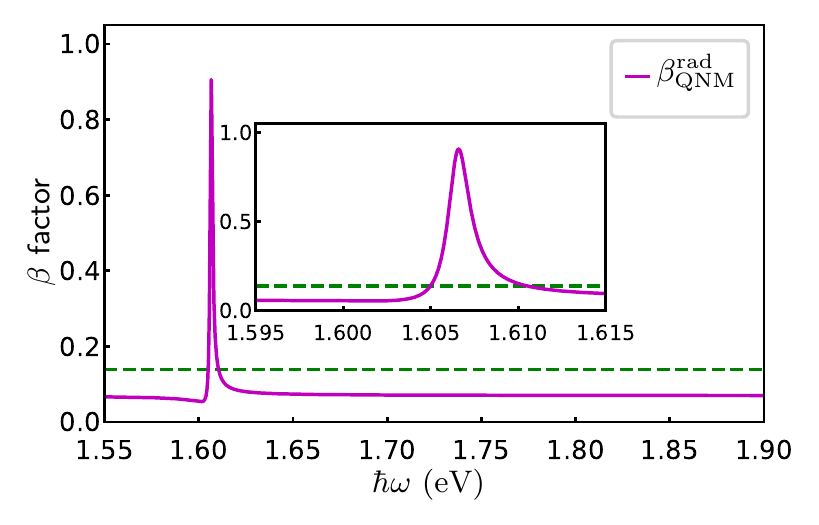}
	\caption{Quantum modal $\beta$-factor  of the metal-dielectric hybrid structure from Fig.~\ref{fig: Scheme}, shown over a broad frequency range for the two-mode quantum QNM model (solid magenta line, Eq.~\eqref{eq: GammaQNMRad}). For the phenomenological dissipative JC model, a phenomenological constant is shown (green dashed line with $\beta\approx 0.14$), which mimics the behaviour of the radiative output at TLS frequencies around $\omega_{\rm pc}$ with respect to the classical $\beta_{\rm class}^{\rm rad}$ solution from Fig.~\ref{fig: beta}. The inset shows a zoom-in around the PC-like mode frequency $\hbar\omega_{\rm pc}=1.6052~({\rm eV)}$. In the QNM-JC model, a peak close to the PC-like mode with frequency $\omega_{\rm pc}$ appears, which is located at the position of the highly non-Lorentz interference effect (dip in $F_p$) from Fig.~\ref{fig: PF_quant_Hybrid}.}  
	\label{fig: betaQuant}
\end{figure}

Next, we discuss the difference between the QNM-JC model and the phenomenological dissipative JC model in terms of (modal) $\beta$-factors. For the QNM-JC model, we define the $\beta$-factor as 
\begin{equation}
    \beta^{\rm rad}_{\rm QNM}=\Gamma^{\rm rad}/\Gamma, 
\end{equation}
where $\Gamma^{\rm rad}$ is the radiative part of the cavity-enhanced spontaneous emission rate, defined through
\begin{equation}
    \Gamma^{\rm rad}=\sum_{\mu,\eta={\rm pc,pl}}S_{\mu\eta}^{\rm rad}\tilde{g}_\mu \tilde{g}_\eta^*\frac{i(\omega_\mu-\omega_\eta)+\gamma_\mu+\gamma_\eta}{(\Delta_{\mu a}-i\gamma_\mu)(\Delta_{\eta a}+i\gamma_\eta)} .\label{eq: GammaQNMRad}
\end{equation}
As one can see from Fig.~\ref{fig: betaQuant}, $\beta^{\rm rad}_{\rm QNM}$ reproduces the overall shape of the full Maxwell solution from Fig.~\ref{fig: beta} (third row) through the presence of $S_{\mu\eta}^{\rm rad}$ and $S_{\mu\eta}^{\rm nrad}$. In contrast, in a phenomenological dissipative JC model, it is not clear at all how to separate radiative and non-radiative contribution, and this model is not even able to predict a frequency-dependent modal $\beta$-factor: While adding a constant $\beta$-factor phenomenologically can yield appropriate results in the single-mode limit, this is non-trivial for more then one mode, as is shown by the (green dashed) line in Fig.~\ref{fig: betaQuant}, since off-diagonal effects between the modes can alter the output behaviour. 

\subsection{Strong light-exciton coupling: quantized system properties\label{subsec: sysprop}}

Next, we turn to the case of the strong QNM-emitter coupling regime; this regime is realized by the parameters shown in Tab.~\ref{Tab: Params} with a dipole moment of $d_\mathrm{a}=10~D$, leading to $|g_{\rm pc}|\approx 8\Gamma_{\rm pc} $, $|g_{\rm pl}|\approx 2\Gamma_{\rm pl} $ and  $|\tilde{g}_{\rm pc}|\approx 8\gamma_{\rm pc} $, $|\tilde{g}_{\rm pl}|\approx 2\gamma_{\rm pl} $. See App.~\ref{app: CouplingRegime} for further discussion on the light-matter coupling regimes. Here,  we use the full master equation (Eq.~(\ref{eq: MasterTwoModeQNMJC}-\ref{eq: MasterTwoModeJC})) beyond the bad cavity limit ($\Gamma_{\rm pc/pl}\gg g_{\rm pc,pl}$ and $\gamma_{\rm pc/pl}\gg \tilde{g}_{\rm pc,pl}$). For the following calculations, we choose the TLS frequency resonant to the PC-like eigenfrequency of both photon Hamiltonians ($H_{\rm em}^{\rm QNM}$ and $\tilde{H}_{\rm em}^{\rm JC}$), i.e., $\hbar\omega_{\rm a} = \epsilon^{(\rm pc)}_{\rm em}$. In the phenomenological dissipative JC model, the PC-like eigenfrequency of $\tilde{H}_{\rm em}^{\rm JC}$ is simply the PC mode frequency itself, i.e. $\epsilon^{(\rm pc)}_{\rm em}=\omega_{\rm pc}$. In contrast, in the QNM-JC model, the PC-related eigenfrequency of $H_{\rm em}^{\rm QNM}$ is slightly red shifted compared to $\omega_{\rm pc}$ (cf. Fig~\ref{fig: PF_quant_Hybrid}) and which we chose as the TLS frequency in the QNM-JC model.

\subsubsection{Eigenenergies and eigenstates}
To provide the essential background for the interpretation of our full master equation results,  we first discuss the eigenenergies and eigenstates of the coupled TLS-QNM system with respect to the QNM-JC  
model and the phenomenological, dissipative JC model without an external pump, i.e., we set $\Omega = 0$ in the Hamiltonians, Eq.~\eqref{eq: HsysTwoModeQNMJC} and \eqref{eq: HsysTwoModeJC}. 
Note that the respective Hamiltonians without external pump are denoted as $H^{\rm QNM}_{\rm sys},H^{\rm JC}_{\rm sys}$.
 \begin{figure}[!h]
	\centering
	\includegraphics[trim={0.20cm 0 0.15cm 0},clip,width=0.99\columnwidth]{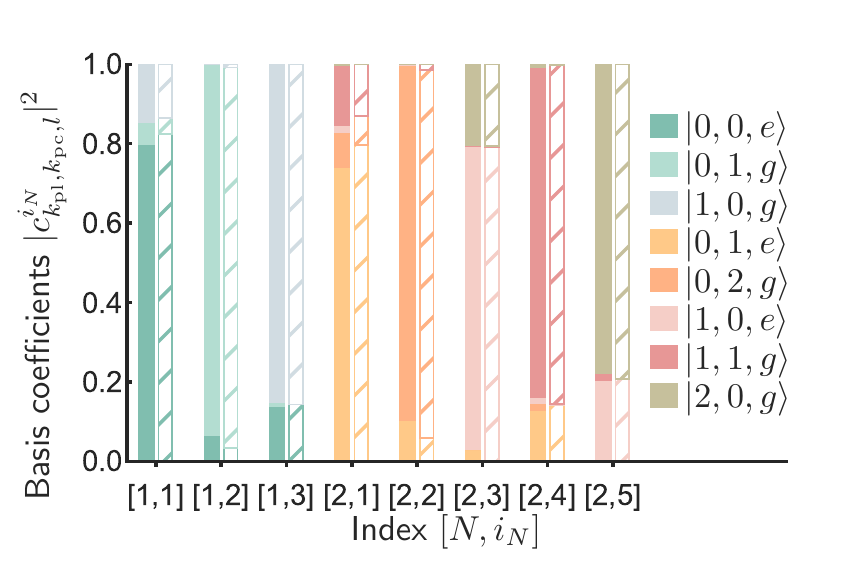}
	\caption{Absolute square of the basis coefficients $c_{k_{\rm pc},k_{\rm pl},l}^{i_N}$ for the one- and two-excitation manifold ($N=1,2$) with the TLS frequency $\omega_{\rm a} = \epsilon_{\rm em}^{(2)}$ aligned with the eigenvalue  of the electromagnetic Hamiltonian $H_{\rm em}$ close to the PC frequency. The solid (QNM-JC model) and dashed (phenomenological dissipative JC model) bars reflect the contribution of the (bare) eigenstates $|k_{\rm pl},k_{\rm pc}, l\rangle$ of the uncoupled photon-exciton system to the eigenstates $|\varphi_{N,i_N}\rangle$ of the actual coupled photon-exciton system. The eigenstates of the coupled system are similar to the respective bare states, since the coupling constants indicate  a regime below  ultrastrong coupling (cf. Tab.~\ref{Tab: Params}). 
	}\label{fig: StateCoeff}
\end{figure}
To obtain a formal inside into the eigenstates of $H_{\rm sys}^{\rm QNM}$ ($H_{\rm sys}^{\rm JC}$), we recall the bare state basis, in which the density operator is expanded, i.e. $|k_{\rm pl},k_{\rm pc},l\rangle$ for the QNM-JC model and $|\tilde{k}_{\rm pl},\tilde{k}_{\rm pc},l\rangle$ for the phenomenological dissipative JC model. Here, $|k_{\rm pl}\rangle(|\tilde{k}_{\rm pl}\rangle)$ represents the number state of the plasmon-like mode, $|k_{\rm pc}\rangle(|\tilde{k}_{\rm pc}\rangle)$ is the number state of the PC-like mode and $l=g,e$ denotes the state of the TLS in the QNM-JC (phenomenological dissipative JC) model. 

These bare states are eigenstates of the photon number operators $\hat{n}_{\rm pl}=A_{\rm pl}^\dagger A_{\rm pl}$ ($\hat{\tilde{n}}_{\rm pl}=\tilde{\alpha}_{\rm pl}^\dagger\tilde{\alpha}_{\rm pl}$),  $\hat{n}_{\rm pc}= A_{\rm pc}^\dagger A_{\rm pc}$ ($\hat{\tilde{n}}_{\rm pc}=\tilde{\alpha}_{\rm pc}^\dagger\tilde{\alpha}_{\rm pc}$) and the TLS number operator $\hat{n}_{\rm a}=\sigma^+ \sigma^-$ (corresponding to the occupation of the upper level $|e\rangle$) with non-degenerate eigenvalues, respectively. However, they are not eigenstates of the Hamiltonians $H^{\rm QNM}_{\rm sys}$ ($H^{\rm JC}_{\rm sys}$), since the subsystems are coupled. The total number operator, e.g. for the QNM-JC model, $\hat{N}=\hat{n}_{\rm pl}+\hat{n}_{\rm pc}+\hat{n}_{\rm a}$ constitutes of the $2N+1$-dimensional eigenspace
\begin{equation}
    \left\{|\phi_{N}\rangle\right\} = {\rm span}\left\{|k_{\rm pl},k_{\rm pc},l\rangle \delta_{N,k_{\rm pl}+k_{\rm pc}+l}\right\}.\label{eq: eigenspaceN}
\end{equation}

Since the total number operator commutes with the system Hamiltonians without external pumping, i.e. $[\hat{N},H^{\rm QNM}_{\rm sys}]=0$, there exists a common eigenbasis of $\hat{N}$ and $H^{\rm QNM}_{\rm sys}$. 
However, since the degeneracy of the eigenvalues of $H^{\rm QNM}_{\rm sys}$ is generally different compared to $\hat{N}$, the eigenbasis formed by the eigenspaces from Eq.~\eqref{eq: eigenspaceN} is not necessarily a eigenbasis of $H^{\rm QNM}_{\rm sys}$. A common eigenbasis with eigenstates $|\varphi_{N,i_N}\rangle$, where $i_N=1,\dots 2N+1$, can be defined as a linear combination of basis elements $|\phi_{N,i_N}\rangle$ of the eigenspaces from Eq.~\eqref{eq: eigenspaceN} with respect to $i_N$. Without loss of generality, we choose $|k_{\rm pl},k_{\rm pc},l;j_N\rangle \delta_{N,k_{\rm pc}+k_{\rm pl}+l}$ with $j_N=1,\dots 2N+1$ as the basis set of $\left\{|\phi_{N}\rangle\right\}$ to construct the states
\begin{equation}
    |\varphi_{N,i_N}\rangle = \sum_{j_N}c_{j_N}^{i_N}|k_{\rm pl},k_{\rm pc},l;j_N\rangle \delta_{N,k_{\rm pc}+k_{\rm pl}+l},
    \end{equation}
for the $N$-th rung in the QNM-JC ladder (and formally equal for the phenomenological dissipative JC model). The states $|\varphi_{N,i_N} \rangle$ are solutions to the eigenvalue problem
\begin{equation}
    H_{\rm sys}^{\rm QNM}|\varphi_{N,i_N} \rangle=E_{N,i_N}|\varphi_{N,i_N} \rangle,
\end{equation}
where $E_{N,i_N}$ are the (real) eigenenergies of the system Hamiltonian $H_{\rm sys}^{\rm QNM}$. Note that we order $i_N$ for a specific manifold $N$, such that $E_{N,i_N}<E_{N,i_N+1}$ for all $i_N$. Although $|\varphi_{N,i_N} \rangle$ are not eigenstates of the full Liouvillians (Eq.~\eqref{eq: MasterTwoModeQNMJC}), they still reflect the effect of coupling between the subsystems and, in contast to the full eigenstates, constitute a orthonormal basis, which will later be used for the definition of a projection operator.

Since the parameters of the hybrid system (cf. Tab.~\ref{Tab: Params}) indicate a regime below the ultrastrong~\cite{frisk_kockum_ultrastrong_2019,RevModPhys.91.025005} (cf. subsection~\ref{Subsec: GFquant}) 
coupling regime ($g_{\rm pc/pl}\ll \omega_{\rm pl,pc}$), each $|\varphi_{N,i_N} \rangle$ is approximately dominated by a single bare state $|n_{\rm pl},n_{\rm pc},l\rangle$, as shown for $N=1$ and $N=2$ in Fig.~\ref{fig: StateCoeff}. It is therefore instructive to rename the eigenstates corresponding to the one-excitation manifold ($N=1$) as $|\varphi_{1,1}\rangle=|\varphi_{\rm a}\rangle$ (similar to the upper state $|e\rangle$), $|\varphi_{1,2}\rangle=|\varphi_{\rm pc}\rangle$ and $|\varphi_{1,3}\rangle=|\varphi_{\rm pl}\rangle$, and the eigenstates corresponding to the two-excitation manifold ($N=2$) as $|\varphi_{2,1}\rangle=|\varphi_{a-\rm pc}\rangle$, $|\varphi_{2,2}\rangle=|\varphi_{\rm pc-pc}\rangle$, $|\varphi_{2,3}\rangle=|\varphi_{a-\rm pl}\rangle$, $|\varphi_{2,4}\rangle=|\varphi_{\rm pc-\rm pl}\rangle$ and $|\varphi_{2,5}\rangle=|\varphi_{\rm pl-\rm pl}\rangle$, respectively. Obviously, the vacuum state remains the same, i.e., $|\varphi_{0,1}\rangle=|0,0,g\rangle=|\varphi_{\rm vac}\rangle$.

\begin{figure}[!h]
	\centering
	\includegraphics[trim={0.20cm 0 0.15cm 0},clip,width=0.99\columnwidth]{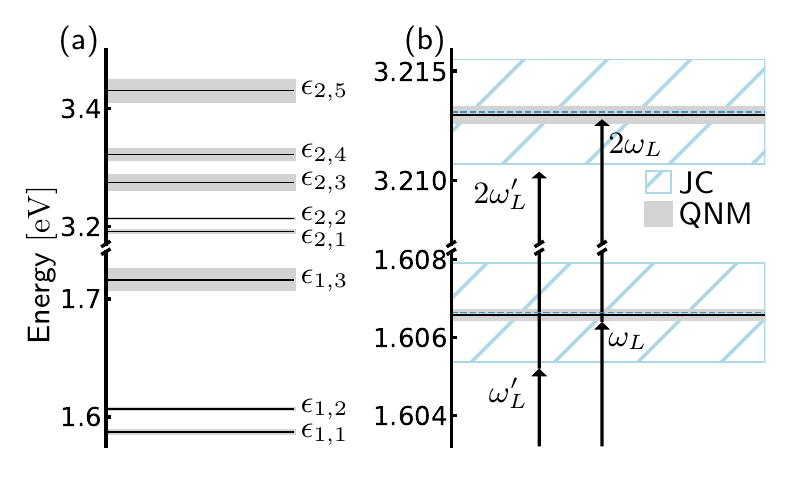}
	\caption{(a) First (3-fold) and second (5-fold) rung of the QNM-JC  energy ladder with $\omega_{\rm a}$ aligned to the eigenfrequency of the electromagnetic Hamiltonian $H_{\rm em}$ close to the PC mode frequency: The solid lines show the real part  of the complex eigenenergies $(\epsilon_{N,i_N})$, while the grey area covers the imaginary part and is bounded by ${\rm Re}(\epsilon_{N,i_N})\pm 0.2{\rm Im}(\epsilon_{N,i_N})$. (b) Eigenenergies $\epsilon_{1,2}$ and $\epsilon_{2,2}$: The solid (lightblue dashed) line shows the real part of the eigenenergies in the QNM-JC model (phenomenological dissipative JC model), while the grey (lightblue dashed) area reflects the imaginary part and is bounded by ${\rm Re}(\epsilon_{N,i_N})\pm {\rm Im}(\epsilon_{N,i_N})$.  Direct 2-photon transitions via an external laser with frequency $\omega_{\rm L} (\omega_{\rm L}')$ are sketched. Note, that in both subfigures, there is a sudden jump on the energy axis from the first to second rung and that the scaling is different in the upper and lower half. 
	}\label{fig: EigEnergies}
\end{figure}

The energies $E_{N,i_N}$ are not eigenenergies of the full Liouvillians (from Eq.~\eqref{eq: MasterTwoModeQNMJC},\eqref{eq: MasterTwoModeJC}), since the additional dissipative terms add imaginary parts to $E_{N,i_N}$ and shifts the real parts due to coupling of the subsystems in the coherent part.  However, the complex eigenenergies $\epsilon_{N,i_N}$ of the non-hermitian Hamiltonian
\begin{equation}
    H_{\rm diss}^{\rm QNM} = H_{\rm sys}^{\rm QNM}-i\Gamma_{\rm pl}A_{\rm pl}^\dagger A_{\rm pl}-i\Gamma_{\rm pc}A_{\rm pc}^\dagger A_{\rm pc}-i\frac{\gamma_{\rm SE}}{2}\sigma^+\sigma^-,\label{eq: Hdiss}
\end{equation}
yield a subset of the eigenenergies of the full Liouvillian $\mathcal{L}^{\rm QNM}=H_{\rm sys}^{\rm QNM}+\mathcal{L}^{\rm QNM}_{\rm diss}$ (cf. Eq.~\eqref{eq: MasterTwoModeQNMJC}), excluding the transition energies~\cite{torres2014closed}. This holds also true in the case of the phenomenological dissipative JC model, where, obviously, $H_{\rm sys}^{\rm QNM}\rightarrow \tilde{H}_{\rm sys}^{\rm JC}$ and $\Gamma_{\rm pl(pc)}\rightarrow\gamma_{\rm pl(pc)}$ in Eq.~\eqref{eq: Hdiss}.

 The complex eigenenergies $\epsilon_{N,i_N}$ for the first and second rung are depicted in Fig.~\ref{fig: EigEnergies}. We note, that the real part of these energies are very similar in the QNM-JC model and phenomenological dissipative JC model, although the full energies for the latter are not shown. However, the imaginary part of the eigenenergies, corresponding to the states dominated by the PC contributions, are very different (cf. Fig.~\ref{fig: EigEnergies}, b), as is the case for the bare rates $\Gamma_{\rm pc}$ and $\gamma_{\rm pc}$ (cf. Tab.~\ref{Tab: Params}). We emphasize that this is a consequence of the symmetrization of the QNM operators in the QNM-JC model (necessary to construct Fock states), which yields symmetrized mode parameters as linear combination of the input QNM parameters, used by the phenomenological dissipative JC model. Furthermore, the increase of the imaginary part in higher rungs ($N>1$) is also different in both models: While in the QNM-JC model, we calculate ${\rm Im}(\epsilon_{\rm pc-pc})\approx 3{\rm Im}(\epsilon_{\rm pc})$, for the phenomenological dissipative JC model, we find ${\rm Im}(\epsilon_{\rm pc-pc})\approx 2{\rm Im}(\epsilon_{\rm pc})$, which can 
 lead to major differences between both models with respect to the response to an external laser in the higher rungs of the energy ladder.

 We note  that, in the following, we will also adopt the notation introduced above for the eigenstates of $H_{\rm sys}^{\rm QNM}$ ($H_{\rm sys}^{\rm JC}$) for $N=1,2$, to the complex eigenenergies, e.g., $\epsilon_{1,1}=\epsilon_{\rm a}$. We also emphasize that the non-Hermitian Hamiltonian (Eq.~\eqref{eq: Hdiss}) is only used as a basis to obtain the complex eigenenergies of the open system to get a first intuitive understanding of the system, since the non-Hermitian Hamiltonian is only valid for short times even in the single excitation manifold, since it misses the quantum jump terms. However, the following simulations of the probabilities and correlation functions are all done with the full master equations (Eqs.~\eqref{eq: MasterTwoModeQNMJC}, \eqref{eq: MasterTwoModeJC}) using the derived Lindbladian for dissipation and not a non-Hermitian Hamiltonian alone.

\subsubsection{Steady-state probabilities and occupation numbers}
Having discussed the eigenenergies and eigenstates of the QNM-JC model and the phenomenological dissipative JC model without pump, we now apply an external optical driving on the system, and simulate the full master equations (Eq.~(\ref{eq: MasterTwoModeQNMJC}-\ref{eq: MasterTwoModeJC})). In this situation, via external pumping, few photon effects can be studied. We choose a Rabi frequency of $\Omega_{\rm L}=0.025|\tilde{g}_{\rm pl}|\sim|g_{\rm pc}|$, which
is in a excitation regime, where effects from the two-excitation manifold of the JC ladder are visible 
(cf. App.~\ref{app: ExcitationRegime} for discussion on excitation regimes) 
and we note, that $\tilde{g}_{\rm pl}$ is the TLS-plasmon coupling constant in the original QNM basis using $\tilde{\mathbf{f}}_{\rm pl}$.

Evaluating Eqs.~\eqref{eq: MasterTwoModeQNMJC} and \eqref{eq: MasterTwoModeJC} numerically, we now analyse the (total) probability to find the system in a 0, 1 or 2-exitation (photon) state as well as the occupation numbers of the two QNMs and the TLS for $t\rightarrow\infty$ (steady-state regime) as a function of laser frequency $\omega_{\rm L}$, so that
we access the intrinsic quantum anharmonicities of the  higher rungs of the JC ladder. The quantities of interest are the occupation numbers $n_{\rm pl}\equiv\langle A_{\rm pl}^\dagger A_{\rm pl}\rangle$,  $n_{\rm pc}\equiv\langle A_{\rm pc}^\dagger A_{\rm pc}\rangle$ and $n_{\rm a}\equiv\langle \sigma^+ \sigma^-\rangle$ for the plasmon-like mode, PC-like mode and TLS upper level $\ket{e}$, respectively, as well as the probabilities $P_{[N,i]}=\langle  \hat{P}_{[N,i]} \rangle$ connected to the projector
\begin{equation}
    \hat{P}_{[N,i_N]}=|\varphi_{N,i_N} \rangle\langle \varphi_{N,i_N}|\label{Prob_Mode},
\end{equation}
on the $i_N$-th eigenstate $|\varphi_{N,i_N} \rangle$ of $H_{\rm sys}^{\rm QNM}$ (or $H_{\rm sys}^{\rm JC}$ for the phenomenological dissipative JC model). For $N=1,2$, we also adopt the notation introduced in the last subsection for the probabilities, e.g., $P_{[1,1]}=P_{\rm a}$. 

(i) \textit{One-excitation manifold}.---First, we study the $1$-excitation manifold as well as the occupation numbers and choose a laser frequency regime around the PC mode frequency $\omega_{\rm pc}$, where the most striking differences between the QNM-JC and the phenomenological dissipative JC model are visible. As explained in the last subsection, the $0$-excitation manifold only contains the trivial vacuum state $|\varphi_{\rm vac}\rangle$ and the $1$-excitation manifold 
yields the three eigenstates $|\varphi_{\rm a}\rangle$, $|\varphi_{\rm pc}\rangle$ and $|\varphi_{\rm pl}\rangle$.  

\begin{figure}[!h]
	\centering
	\includegraphics[trim={0.20cm 0 0.15cm 0},clip]{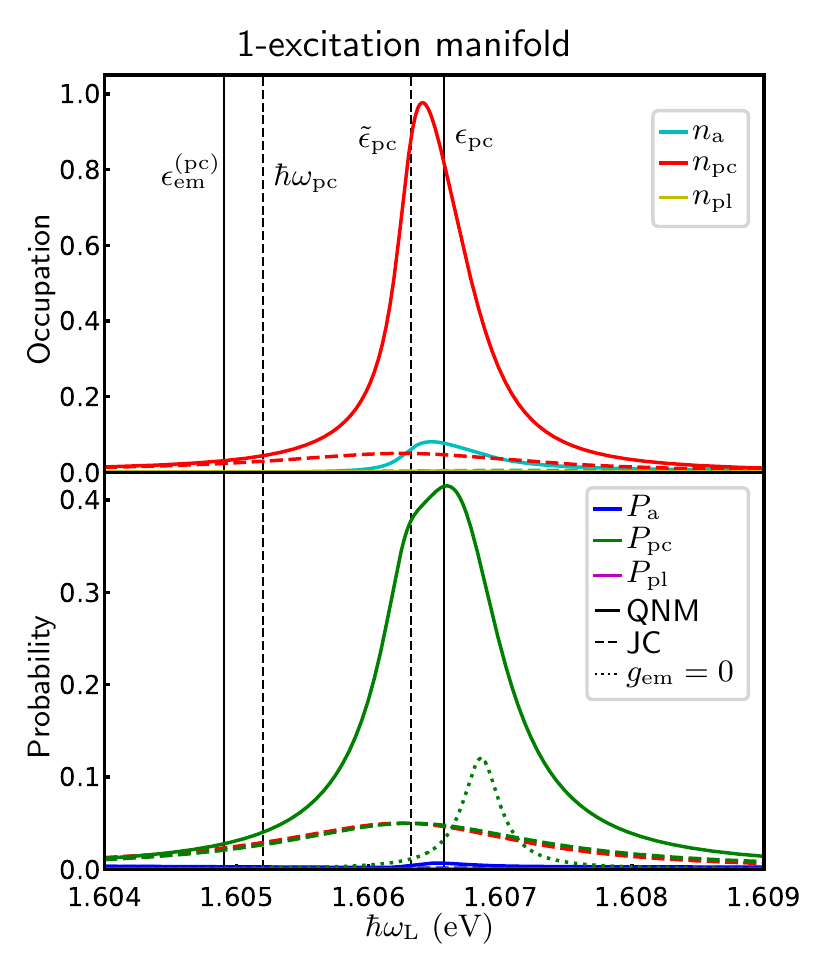}
	\caption{Occupation numbers $n_{\rm a},n_{\rm pc},n_{\rm pl}$ and probabilities $P_{\rm a}, P_{\rm pc},P_{\rm pl}$ 
	of the TLS (corresponding to upper level $|e\rangle$), plasmon and PC-like mode in the steady state, obtained with the QNM-JC model (solid) and phenomenological dissipative JC model (dashed) as functions of laser frequency $\omega_{\rm L}$ around $\omega_{\rm pc}$. The TLS frequency $\hbar\omega_{\rm a} = \epsilon^{(2)}_{\rm em}$ is aligned to the eigenenergy of the full electromagnetic Hamiltonian $H_{\rm em}$ ($\omega_{\rm pc}$ in the phenomenological dissipative JC model) and  the TLS is pumped with an external laser with Rabi frequency $\Omega\sim |g_{\rm pc}|$, where $|g_{\rm pc}|\approx 8\Gamma_{\rm pc}$ is the TLS-PC coupling constant (in the strong photon-exciton coupling regime) and $\Gamma_{\rm pc}$ is the PC mode decay rate (cf. Tab.~\ref{Tab: Params}). Additionally, we show $P_{\rm pc}$ for the QNM-JC model in the limit $g_{\rm em}=0$ (dotted line), cf. Eq.~\eqref{eq: HsysTwoModeQNMJC}. 
	}\label{fig: Fock_All1}
\end{figure}

The results are shown in Fig.~\ref{fig: Fock_All1},  and we start with the analysis of $P_{\rm a}$ and $n_{\rm a}$, corresponding to the probability and occupation of the TLS-like upper level, respectively. The peak heights of both quantities differ strongly in the two different models: Whereas in the phenomenological dissipative JC model, $n_{\rm a}<10^{-2}$, in the QNM-JC model $n_{\rm a}\approx 0.1$, i.e. one order of
magnitude larger. This is because of the stronger coupling ($|\tilde{g}_{\rm pc}|\approx 10|\tilde{g}_{\rm pc}|$) between the TLS and the PC like-mode and the larger depopulation rates ($\gamma_{\rm pc}\approx 10\Gamma_{\rm pc}$) of the PC-like states in the phenomenological dissipative JC model. Quantitatively, this difference is similar for $P_{\rm a}$; however, $P_{\rm a}$ itself is much smaller than  $n_{\rm a}$, which is consequence of $|\varphi_{\rm a}\rangle$ being a linear combination of the bare states $|1,0,g\rangle$, $|0,1,g\rangle$ and $|0,0,e\rangle$, which leads to  
a reduction of the diagonal contribution $|0,0,e\rangle\langle 0,0,e|$ (cf. Fig.~\ref{fig: StateCoeff}) in $P_{\rm a}$.

Next we look at $P_{\rm pl}$ and $n_{\rm pl}$, corresponding to the probability and occupation number of the plasmon-like mode, respectively. The probability $P_{\rm pl}$  has a negligible value, $P_{\rm pl}<10^{-3}$, for all values of  $\omega_{\rm L}$ that we tried, which is the case for both models. Furthermore, the occupation number $n_{\rm pl}$ of the plasmon-like mode is very small $(<10^{-2})$ and also very similar in both models. The similarity for the plasmon-like mode is a consequence of the small differences in the plasmon decay rate ($\Gamma_{\rm pl}\approx\gamma_{\rm pl}$) and plasmon-TLS coupling  constant ($|g_{\rm pl}|\approx|\tilde{g}_{\rm pl}|$) before and after symmetrization and diagonalization (see also Tab.~\ref{Tab: Params}).
The small values are a consequence of the high plasmon decay rates $\Gamma_{\rm pl},\gamma_{\rm pl}$, which results in a fast depopulation of the plasmon-like states.

The most pronounced changes appear in $P_{\rm pc}$ and $n_{\rm pc}$ of the PC-like mode. 
First, the peak of $P_{\rm pc}$ in the QNM-JC model is slightly detuned to higher frequencies compared to $\tilde{P}_{\rm pc}$ of the phenomenological dissipative JC model. This is a consequence of the small shift of the PC-like frequency ($\omega_{\rm pc}$) after symmetrization and diagonalization ($\Omega_{\rm pc}$). Second, the spectral width of the probability and occupation number dynamics is much broader in the case of the phenomenological dissipative JC model. This is again a consequence of the width $\gamma_{\rm pc}$ of the PC-related resonance of the phenomenological dissipative JC model being about one order of magnitude broader compared to the effective PC width $\Gamma_{\rm pc}$ of the QNM-JC model ($\gamma_{\rm pc}\approx 10\Gamma_{\rm pc}$). Therefore, the laser can effectively excite a much broader range of the first rung in the phenomenological dissipative JC model ladder.
Third, the peak height of $P_{\rm pc}$ and $n_{\rm pc}$ between both models is completely different; $P_{\rm pc}$ is about one order of magnitude larger than $\tilde{P}_{\rm pc}$, which means that the system has a 10 times larger probability to be in the state  $|\varphi_{{\rm pc}} \rangle$ in the QNM-JC model. Of course, this is because of the fast depopulation of the first rung in the phenomenological dissipative JC model due to the large decay rates. Due to the same reason, there is also a pronounced difference in the peak height of $n_{\rm pc}$. 

There are two further interesting observations that are connected to $n_{\rm pc}$  and $\tilde{n}_{\rm pc}$. First, in the phenomenological dissipative JC model, $\tilde{n}_{\rm pc}$ is nearly identical to $\tilde{P}_{\rm pc}$ close to $\omega_{\rm pc}$, as shown in Fig.~\ref{fig: Fock_All1} (bottom, red and green dashed curve). To explain this, we recall that 
\begin{equation}
  n_{\rm pc}=\sum_{k_{\rm pl}, k_{\rm pc},i}  k_{\rm pc}\langle|k_{\rm pc},k_{\rm pc},i\rangle\langle k_{\rm pl},k_{\rm pc},i|\rangle,
\end{equation}
and in the phenomenological dissipative JC model, we observe that $n_{\rm pc}\approx \langle|0,1,g\rangle\langle 0,1,g|\rangle\approx P_{\rm pc} $. In contrast, in the QNM JC-model, $n_{\rm pc}$ is different to $P_{\rm pc}$ with respect to height and peak position, indicating that higher rung ($N>1$) probabilities are also important here, as we will show below. 
Second, an additional indicator for processes on higher rungs in the QNM-JC model is a start of a spectral hole burning process at the peak of $P_{\rm pc}$, which comes from higher photon probabilities with a smaller laser excitation width.

\begin{figure}[!h]
	\centering
	\includegraphics[trim={0.20cm 0 0.15cm 0},clip]{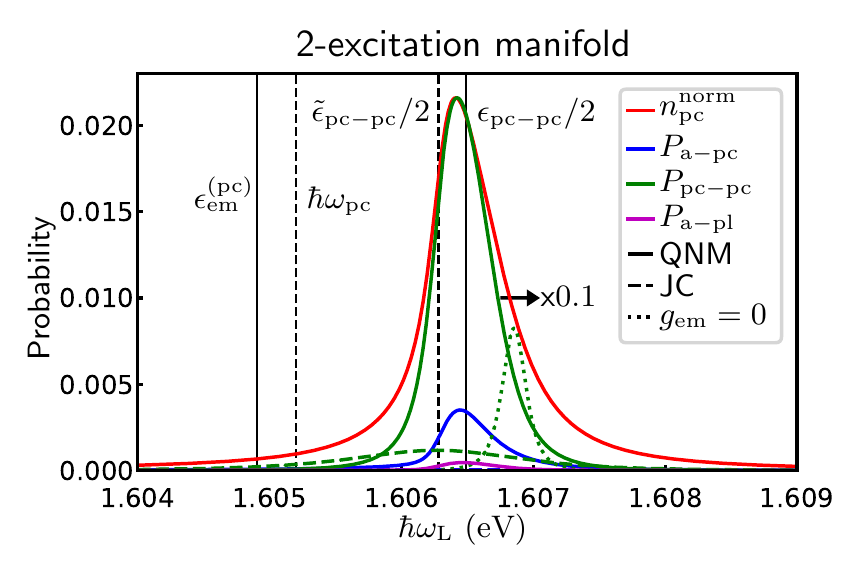}
	\caption{Probabilities $P_{a-\rm pc}, P_{\rm pc-pc}, P_{a-\rm pl}$ corresponding to the $2$-excitation manifold in the steady state obtained with the QNM-JC model (solid) and phenomenological dissipative JC model (dashed) as functions of laser frequency $\omega_{\rm L}$ around the PC mode resonance $\omega_{\rm pc}$. The TLS frequency $\hbar\omega_{\rm a} = \epsilon^{(\rm pc)}_{\rm em}$ is aligned to the eigenenergy of the full electromagentic Hamiltonian $H_{\rm em}$ ($\omega_{\rm pc}$ in the phenomenological dissipative JC model) and the TLS is pumped with an external laser with Rabi frequency $\Omega\sim |g_{\rm pc}|$, where $|g_{\rm pc}|\approx 8\Gamma_{\rm pc}$ is the TLS-PC coupling constant (in the strong photon-exciton coupling regime) and $\Gamma_{\rm pc}$ is the PC mode decay rate (cf. Tab.~\ref{Tab: Params}). We also show $P_{\rm pc-pc}$ for the QNM-JC model in the limit $g_{\rm em}=0$ (dotted line), cf. Eq.~\eqref{eq: HsysTwoModeQNMJC}, and the occupation number $n_{\rm pc}$ normalized to the same height as $0.1P_{\rm pc-pc}$. Note, that all other contributions corresponding to the two-excitation manifold are negligible and not shown here. 
	}\label{fig: Fock_All2}
\end{figure}

(ii) \textit{Two-excitation manifold}.---Next, we discuss the probabilities as a function of laser frequency around the PC mode frequency, connected to the 2-excitation manifold of the QNM-JC and phenomenological dissipative JC ladder, which consists of the five eigenstates: 
 $|\varphi_{a-{\rm pc}} \rangle$, $|\varphi_{\rm pc-pc} \rangle$, $|\varphi_{a-{\rm pl}} \rangle$, $|\varphi_{\rm pc-pl}\rangle$ and $|\varphi_{\rm pl-pl} \rangle$. 
The dominant probabilities with peak heights $P>0.001$ are plotted in Fig.~\ref{fig: Fock_All2} ($P_{a-\rm pc}, P_{\rm pc-pc}, P_{a-\rm pl}$). All other contributions ($P_{\rm pl - pc}, P_{\rm pl-pl}$) are not shown. We notice, that quantitative differences of $P_{a-\rm pc}, P_{a-\rm pl}$ between both models behave similar to the differences in $P_{\rm a}$ from the 1-excitation manifold. However, in both models, these higher-rung contributions are negligible compared to the 1-excitation manifold contributions. Thus, below we concentrate on $P_{\rm pc-pc}$, which covers the most striking and interesting differences.

We observe that $P_{\rm pc-pc}$ is roughly two order of magnitude higher compared to $\tilde{P}_{\rm pc-pc}$, which constitutes a much larger difference compared to the case of the one-excitation manifold. This leads to a different ratio of the maximum values $R_{1-2}={\rm max}\left(P_{1,i}\right)/{\rm max}\left(P_{2,i}\right)$ between the one- and two-excitation regime: In the QNM-JC model, we obtain $R_{1-2}\approx 2$, while in the phenomenological dissipative JC model, we obtain $R_{1-2}\approx 10^2$.
To help explain this, we show
the two photon resonance ($2\omega_{\rm L}$) close to the peak of the 2-photon probability $P_{\rm PC-PC}$ of the QNM-JC model in the eigenenergy diagram from Fig.~\ref{fig: EigEnergies} (b), showing the eigenstates $|\varphi_{\rm pc} \rangle$ and $|\varphi_{\rm PC-PC} \rangle$: In the QNM-JC model, $\omega_{\rm L}$ at the peak of $P_{\rm PC-PC}$ is not in the range of the PC-like eigenenergy ${\rm Re}(\epsilon_{\rm pc})\pm{\rm Im}(\epsilon_{\rm pc})$. However, the two photon resonance ($2\omega_{\rm L}$) is located in the range of the eigenenergy ${\rm Re}(\epsilon_{\rm pc-pc})\pm{\rm Im}(\epsilon_{\rm pc-pc})$ corresponding to PC-like eigenenergy on the 2nd rung. This means, that there is a direct population of the 2nd rung in the QNM-JC model via a virtual state with smaller energy compared to ${\rm Re}(\epsilon_{\rm pc})-{\rm Im}(\epsilon_{\rm pc})$, leading to the relative high 2-photon probability. This is possible due to the anharmonicity of the JC ladder, i.e., the transitions ${\rm Re}(\epsilon_{N+1,i})-{\rm Re}(\epsilon_{N,i})$ between different manifolds $N$ and $N+1$ depends on $N$ due to the emitter-photon coupling. Of course, this anharmonicity is present in both models, the QNM-JC model and the phenomenological dissipative JC model, but takes a different value in the QNM-JC model as a direct consequence of $H_{\rm em}$ not commuting with the photon number operators. However, in the phenomenological model, the widths of the eigenenergies in the one-excitation manifold are much larger than the difference $D_{\rm pc}\equiv |{\rm Re}(\epsilon_{\rm pc-pc})-2{\rm Re}(\epsilon_{\rm pc})|$ of the transitions, and thus, the first rung is already majorly excited at $\omega_{\rm L}>\omega_{\rm L}'$ and the probability of an indirect population of the second rung via a virtual state is very small (cf. Fig.~\ref{fig: EigEnergies}, b). We note, that this effect in the QNM-JC model is stable against a variation of $\omega_{\rm a}$ with respect to the PC-mode frequency.
Additionally, it is worth to note, that the peak position of the occupation number $n_{\rm pc}$ is nearly directly located at the peak of $P_{\rm pc-pc}$ (cf. Fig.~\ref{fig: Fock_All2}).

We briefly summarize the analysis in this subsection: For moderate pumping with respect to the PC-mode, $\Omega\sim g_{\rm pc}$, the QNM-JC model exhibits a relatively high probability to be in a 1 or 2 excitation state for the PC-like mode ($P_{\rm pc}\sim 0.4$ and $P_{\rm pc-pc}\sim 0.2$), while the vacuum state $|\varphi_{\rm vac}\rangle$ is surpressed, when the laser is tuned in the regime of the PC-like cavity mode frequency.
In contrast, in the phenomenological dissipative JC model, the overall behavior of the resonance structure is similar, but qualitative and quantitative differences are present: The 2-photon probability in the phenomenological dissipative JC model is negligible over the inspected laser frequency regime, caused by the larger decay rate $\gamma_{\rm pc}$ of the PC-like mode. In particular, the maximum peak of the 2-photon probability $P_{\rm pc-pc}$ close to the PC-like mode frequency, is about two orders of magnitude smaller compared to the peak of the QNM-JC model (cf. Fig.~\ref{fig: Fock_All2}). This significant difference between both models is caused by the presence of the inter-mode coupling in the Hamiltonian, Eq.~\eqref{eq: HsysTwoModeQNMJC}, (the case $g_{\rm em}\rightarrow 0$ is indicated by the dotted line in Fig.~\ref{fig: Fock_All1} and \ref{fig: Fock_All2} (bottom)) and the shift of the PC decay rate (cf. Eq.~\eqref{eq: EigVal}), but not caused by the TLS-QNM coupling renormalization: While in the phenomenological dissipative JC model the coupling constant $\tilde{g}_{\rm pc}$ of the TLS to the PC-like cavity mode is about one order of magnitude larger (cf. Tab.~\ref{Tab: Params}), the decay rate $\gamma_{\rm pc}$ of the PC-like mode is also one order of magnitude larger, leading to the same ratios of both quantities, i.e. $|g_{\mu}|/(2\Gamma_\mu)\sim |\tilde{g}_{\mu}|/(2\gamma_\mu)$, as discussed earlier.

Therefore, both models are in the strong light-matter coupling regime, and the difference in the response of the system to the external pump is mainly influenced by the different PC decay rates and the intermode coupling Hamiltonian with coupling $g_{\rm em}$. 
Thus, while the system described on the basis of the phenomenological dissipative JC model is (mainly) in the single-photon regime, the system described by the QNM-JC model has significant multiphoton properties. This shows, that the phenomenological introduction of a two-mode parameter set is not trivial in any way, since the input parameter drastically change because of dissipation itself in the course of deriving the QNM master equation, since decay rates and coupling strength are not independent from each other due to symmetrization ($\mathbf{S}$) and diagonalization ($\mathbf{U}^{(-)}$). Therefore phenomenological models can clearly miss important features in the multiphoton regime.

It is important to stress that our QNM quantization model can be used
to predict new regimes in dissipative quantum optics, since all parameters are calculated on a solid foundation through the QNM eigenfunctions and eigenvalues without any phenomenological approaches.

\subsection{Strong light-electron coupling: Output properties\label{subsec: Outputprop}}

Next, we analyse the impact of the off-diagonal QNM coupling on experimental observables, represented by the derived output electric fields, cp. Section~\ref{Subsec: OutputEfield}. 
We focus on correlation functions of the cavity output field, measurable in specific detector setups.  
For this situation, a detector (e.g.,  a lens which collects light over a wide angle) is modelled as an intensity measurement device on a far field surface $\mathcal{S}$.

\subsubsection{Output far field intensity}
\begin{figure}[!h]
	\centering
	\includegraphics[trim={0.20cm 0 0.15cm 0},clip]{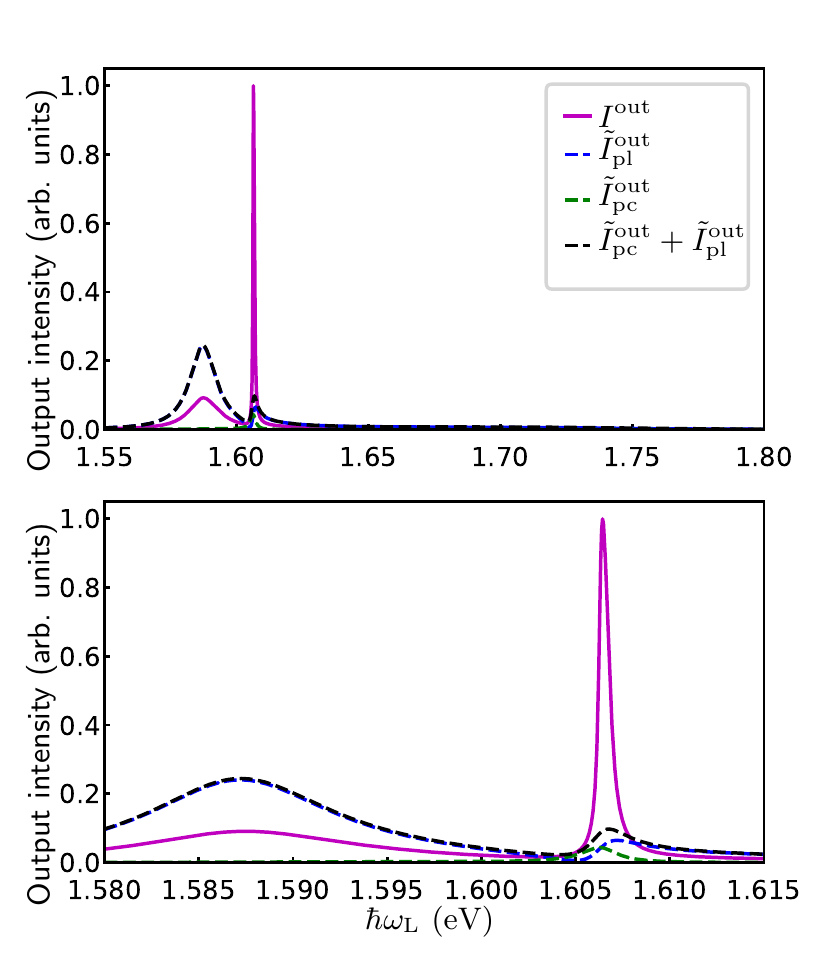}
	\caption{Output intensity function $I^{\rm out}$ measured on a sphere in the far field as function of laser frequency $\omega_{\rm L}$ for the QNM-JC model (solid) and the phenomenological dissipative JC model (dashed) in the steady state with the TLS frequency $\hbar\omega_{\rm a} = \epsilon^{(\rm pc)}_{\rm em}$ aligned to the eigenenergy of the full electromagentic Hamiltonian $H_{\rm em}$. The Rabi frequency of the external pump is $\Omega_{\rm L}\sim|g_{\rm pc}|$, where the TLS-PC coupling is in the strong coupling regime, i.e., $|g_{\rm pc}|\approx 8\Gamma_{\rm pc}$ (cf. Tab.~\ref{Tab: Params}).
	The lower plot shows a zoom-in close to the PC-mode frequency $\omega_{\rm pc}$.
	The quantities are normalized to ${\rm max}(I^{\rm out})$
	from the QNM-JC model. }\label{fig: intensity-out}
\end{figure}

The first quantity of interest is the mean value of the output intensity 
\begin{gather}
    I^{\rm out}\equiv\frac{2\epsilon_0n_{\rm B} c}{\hbar}\sum_i\oint_{\mathcal{S}}\langle E^{(-)}_{{\rm out},i}(\mathbf{s})E^{(+)}_{{\rm out},i}(\mathbf{s})\rangle_{\rm ss},
\end{gather}
where $\mathcal{S}$ is the detector surface and $E^{(+)}_{{\rm out},i}$ is given as in Eq.~\eqref{eq: Eout2} (or Eq.~\eqref{eq: Eout3}). Carrying out the surface integrals by using the definition of the dissipation matrix $S_{\mu\eta}^{\rm rad}$, and by recalling that the input state associated to, e.g., $A_\mu^{\rm in}$, is chosen as the vacuum state, leads to the form (cf. App.~\ref{app: G2outDerivation})
\begin{gather}
    \bar{I}^{\rm out}=\sum_{\mu,\mu'}L_{\mu\mu'}\langle A_\mu^\dagger A_{\mu'}\rangle_{\rm ss}\label{eq: IoutQNMfinal}, 
\end{gather}
where $A_\mu^{(\dagger)}$ represent the annihilation (creation) operator of the symmetrized and diagonalized QNM $\mu=\rm pl,pc$. Furthermore,
\begin{equation}
L_{\mu\eta}=\sum_{\substack{\mu',\mu''\\ \eta',\eta''}}U^{(-)*}_{\mu'\mu}\left(\mathbf{S}^{1/2}\right)_{\mu'\mu''}\tilde{L}_{\mu''\eta''}\left(\mathbf{S}^{1/2}\right)_{\eta''\eta'} U^{(-)}_{\eta'\eta},\label{eq: Lfinal}
\end{equation}
and
\begin{align}
    \tilde{L}_{\mu\eta}
    =&-2i\frac{\omega_{\mu}\omega_{\eta}(\omega_{\mu}-\omega_{\eta}^*)}{\omega_{\mu}+\omega_{\eta}} S^{\rm rad}_{\eta\mu}.\label{eq: Ltildefinal}
\end{align}

Clearly the output coupling matrix $L_{\mu\eta}$ is connected to the radiative part of the QNM decay through the dissipation-induced radiative coupling matrix $S_{\mu\eta}^{\rm rad}$.
In contrast, in a phenomenological dissipative JC output model, one assumes typically uncoupled output radiation, where each mode is coupled out to the surrounding environment independently~\cite{koenderink2017single,rousseaux2018quantum}. In the following, we will compare the above formulas using the derived output electric field with the output intensity in a phenomenological dissipative JC model $\tilde{I}^{\rm out}=\sum_\mu \tilde{I}^{\rm out}_{\mu}$ with
\begin{align}
    \tilde{I}^{\rm out}_{\mu}\equiv2\omega_\mu\beta^{\rm rad}(\omega_{\rm a})\gamma_{\mu}\langle \tilde{\alpha}_\mu^\dagger \tilde{\alpha}_\mu\rangle\label{eq: Imu}, 
\end{align}
for the individual output intensities in the far field  of the QNMs $\mu={\rm pc,\rm pl}$ and in the same units as $I^{\rm out}$. 

Note, that the factor $\beta^{\rm rad}(\omega_{\rm a})$ was added phenomenologiaclly as the radiative $\beta$ factor of the hybrid system at the TLS frequency  (cf. Fig.~\ref{fig: beta}, (circles) and Fig.~\ref{fig: betaQuant}, (dashed line)). 
This is another ambiguity of the phenomenological model, since there is no clear separation between radiative and non-radiative decay processes. This is because in this case, $S_{\mu\eta}$ is approximated as a Kronecker-delta from the beginning and therefore, these $\beta$ factors have to be added phenomenologically. In fact, taking the beta factors in a hybrid structure at certain frequencies is a highly non-trivial choice, since $\beta^{\rm rad}(\omega)$ changes drastically as a function of frequency close to $\omega_{\rm pc}$, as shown in Fig.~\ref{fig: beta} (bottom). This is usually a sign for non-Markovian output characteristics. 
This frequency dependent beta factor is captured (at least in the bad cavity limit, where a comparison to a full Maxwell solution is possible) by the specific form of $S_{\mu\eta}^{\rm rad}$ and $S_{\mu\eta}^{\rm nrad}$ in the QNM-JC model.
To underline the differences, we show the different output intensities in the steady state as function of the laser frequency $\omega_{\rm L}$ in Fig~\ref{fig: intensity-out}.

The peak height of $I^{\rm out}$ in the full QNM-JC model at the PC frequency is roughly one order of magnitude larger than in a phenomenological treatment using the formulas in Eq.~\eqref{eq: Imu}.  Obviously, the output coupling is drastically increased in the full QNM JC-model, when the laser is tuned to the PC frequency regime and results from the increase of the beta factor in the regime close to the PC-like eigenfrequency. {This has a major impact on modelling of hybrid structures for 
nonlinear cavity-QED experiments, since the phenomenological model (even if the system master equation is fitted appropriately) highly underestimates the output coupling.}

\subsubsection{Second-order quantum correlation functions}
\begin{figure}[!h]
	\centering
	\includegraphics[trim={0.20cm 0 0.15cm 0},clip]{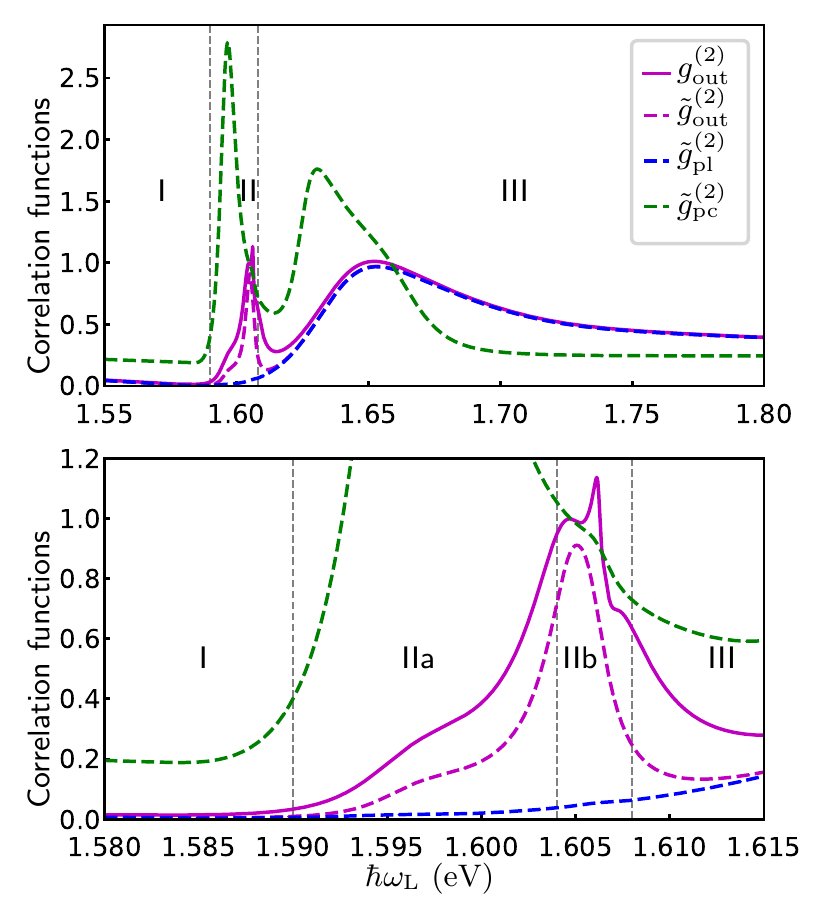}
	\caption{
	Normalized second-order quantum correlation functions in the steady-state regime for the plasmonic-like mode (blue), photonic-crystal mode (green) and the output electric field (magenta) for the QNM-JC model (solid) and the phenomenological dissipative JC model (dashed) over laser frequency $\omega_{\rm L}$.   
	Lower plot shows a zoom-in close to the PC-mode frequency $\omega_{\rm pc}$.  }\label{fig: g2sys-out}
\end{figure}

Next, we turn to the stationary normalized second-order correlation functions $g^{(2)}$. For a detector as a sphere $S$ (as explained above), this correlation function reads in general
\begin{gather}
    g^{(2)}_{\rm out}{\equiv}\frac{\sum_{ij}\oint_{\mathcal{S}}\oint_{\mathcal{S}'}\langle E^{(-)}_{{\rm out},i}(\mathbf{s})E^{(-)}_{{\rm out},j}(\mathbf{s}')E^{(+)}_{{\rm out},j}(\mathbf{s}')E^{(+)}_{{\rm out},i}(\mathbf{s})\rangle_{\rm ss}}{\left[\sum_i\oint_{\mathcal{S}}\langle E^{(-)}_{{\rm out},i}(\mathbf{s})E^{(+)}_{{\rm out},i}(\mathbf{s})\rangle_{\rm ss}\right]^2},
\end{gather}
where again $E^{(+)}_{{\rm out},i}$ is given as in Eq.~\eqref{eq: Eout2} (or Eq.~\eqref{eq: Eout3}). Carrying out the surface integrals similar to $I^{\rm out}$ 
leads to the form (cf. App.~\ref{app: G2outDerivation})
\begin{gather}
    g^{(2)}_{\rm out}=\frac{\sum_{\mu',\mu',\eta,\eta'}L_{\mu\mu'}L_{\eta\eta'}\langle A_\mu^\dagger A_\eta^\dagger A_{\eta'} A_{\mu'}\rangle_{\rm ss}}{\left[\sum_{\mu',\mu'}L_{\mu\mu'}\langle A_\mu^\dagger A_{\mu'}\rangle_{\rm ss}\right]^2}\label{eq: g2Final},
\end{gather}
and $L_{\mu\mu'}$ is given in Eq.~\eqref{eq: Lfinal}.

For the phenomenological dissipative JC model, we use the phenomenological second-order correlation output function
\begin{gather}
    \tilde{g}^{(2)}_{\rm out}\equiv\frac{\sum_{\mu,\eta} \omega_\mu\omega_\eta\gamma_\mu\gamma_\eta\langle \tilde{\alpha}_\mu^\dagger \tilde{\alpha}_\eta^\dagger \tilde{\alpha}_\eta \tilde{\alpha}_\mu\rangle_{\rm ss}}{\left[\sum_\mu \omega_\mu\gamma_\mu\langle \tilde{\alpha}_\mu^\dagger \tilde{\alpha}_\mu\rangle_{\rm ss}\right]^2}\label{eq: g2JC}, 
\end{gather}
which is in line with the assumptions used to obtain $\tilde{I}^{\rm out}$ from Eq.~\eqref{eq: Imu}. However, we note that in field of quantum plasmonics and quantum optics, the second-order correlation functions are often assumed to be
\begin{gather}
    \tilde{g}^{(2)}_{\mu}\equiv\frac{\langle \tilde{\alpha}_\mu^\dagger \tilde{\alpha}_\mu^\dagger \tilde{\alpha}_\mu \tilde{\alpha}_\mu\rangle_{\rm ss}}{\langle \tilde{\alpha}_\mu^\dagger \tilde{\alpha}_\mu\rangle_{\rm ss}}\label{eq: g2mu},
\end{gather}
for the individual modes $\mu={\rm pc,\rm pl}$. We emphasize, that the quantities in Eq.~\eqref{eq: g2mu} are not observables, but can in general be useful to characterize the emitter system properties, and are often used as if they were observables~\cite{majumdar2012loss, zhang2014optimal}.

In Fig.~\ref{fig: g2sys-out}, we show the correlation functions $\tilde{g}^{(2)}_{\rm pl}, \tilde{g}^{(2)}_{\rm pc}$ and the output correlation functions $g^{(2)}_{\rm out}, \tilde{g}^{(2)}_{\rm out}$ as functions of the laser frequency $\omega_{\rm L}$. For off-resonant pumping, i.e., from $\hbar\omega_{\rm L}=1.55~{\rm eV}$ to $\hbar\omega_{\rm L}=1.59~{\rm eV}$ (sector I of Fig.~\ref{fig: g2sys-out}), both the system and the output correlation functions show anti-bunched character $g^{(2)}(0)<1$ of the emitted light. In particular, the output correlation functions $g^{(2)}_{\rm out}$,  $\tilde{g}^{(2)}_{\rm out}$ and the plasmon-like correlation function $g^{(2)}_{\rm pl}$ are nearly identical to each other. This is expected, since the hybrid is dominated in the frequency regime by the plasmonic part of the two fundamental QNMs. However, close to the PC-like resonance of the hybrid, for $\hbar\omega_{\rm L}$ in $[1.59,1.604]~{\rm eV}$ (sector IIb in Fig.~\ref{fig: g2sys-out}), the light statistics is differently predicted as bunched light $g^{(2)}(0)>1$ by the PC-system correlations, and not as anti-bunched as shown by the emitted light through the full output correlation $g^{(2)}_{\rm out}$. Although the phenomenological output function $\tilde{g}^{(2)}_{\rm out}$ shows also anti-bunched light, it still differs slightly from $g^{(2)}_{\rm out}$.  The plasmon-like correlation function shows also anti-bunched character in this laser regime, even one order of magnitude below $g^{(2)}_{\rm out}$. 

Another interesting laser frequency regime is $\hbar\omega_{\rm L}\in [1.604,1.608]~{\rm eV}$ (sector IIb in Fig.~\ref{fig: g2sys-out}); here the phenomenological output function $\tilde{g}^{(2)}_{\rm out}$, Eq.~\eqref{eq: g2JC}, predicts anti-bunched light while the full output correlation is clearly above one, showing bunched light statistics. Furthermore, both, $\tilde{g}^{(2)}_{\rm pl}$ and $\tilde{g}^{(2)}_{\rm pc}$ show anti-bunched character. Thus, only the result from QNM-JC model predicts bunched light in this small but important laser regime. Between $\hbar\omega_{\rm L}=1.608 {\rm eV}$ and $\hbar\omega_{\rm L}=1.8 {\rm eV}$ (sector III in Fig.~\ref{fig: g2sys-out}), the light statistics is again erroneously predicted by the PC-like correlation as bunched light. In addition, the phenomenological output function $\tilde{g}^{(2)}_{\rm out}$ converges much faster to the plasmon-like correlation function $\tilde{g}^{(2)}_{\rm pl}$ here. In a frequency regime towards the plasmon-like frequency, the plasmon system correlations become identical to both output correlations. Therefore, the light statistics has changed  considerably especially in the regime close to the PC-like mode frequency $\omega_{\rm pc}$.

To summarize this subsection, the PC-like correlation function $\tilde{g}^{(2)}_{\rm pc}$ is not reliable to model the second-order output correlations in the 2-mode hybrid system over the whole laser excitation frequency regime under consideration, although it seems to be the dominant part of the TLS response to the electromagnetic field at the chosen TLS frequency 
(cf.~Fig.~\ref{fig: PF_quant_Hybrid}). On the other hand, at a laser frequency regime sufficiently enough away from the PC-mode frequency $\omega_{\rm pc}$, the plasmon-like system correlation function $\tilde{g}^{(2)}_{\rm pc}$ is a very good approximation to the output correlation functions $\tilde{g}^{(2)}_{\rm out}$, $g^{(2)}_{\rm out}$, which is expected, since the main coupling regime of the QNMs is near the PC-mode resonance. {Therefore, we can conclude, that in dissipative resonator structures with at least two fundamental (overlapping) modes, the system correlation functions do not reflect the actual quantum properties of the emitted light (at least in the overlap regime), since the output is formed by a linear combination of the coupled modes, which depends on the dissipation and the radiative and non-radiative properties of the system. Furthermore, even if a $\beta$ factor is added to the formulas of a phenomenological dissipative JC model (Eq.~\eqref{eq: g2JC}, it also cannot reproduce the light statistics correctly, since there is an additional non-diagonal coupling in the output quantities, induced by the off-diagonal mode coupling
through the symmetrization.}

\section{Conclusions\label{Sec: Conlusions}}
We have presented a detailed extension of the QNM quantum model from Ref.~\onlinecite{PhysRevLett.122.213901} by including external pumping and by deriving an generalized input-output theory for multiple QNMs. 
Explicit expressions for correlation functions outside of the nanostructure were provided. Furthermore, pronounced differences in the correlation functions inside the system and at the outside detector were found.
We analysed the cavity-QED behaviour of a metal-dielectric hybrid resonator coupled to a TLS in the strong coupling regime. 
We compared our full microscopic QNM model including mode coupling with a phenomenological dissipative JC model.
Significant qualitative and quantitative differences, induced by the inter-mode coupling, were found between  both models. 

In the nonlinear pumping regime,
we also studied the response of the hybrid 
cavity system to an external laser (in a excitation regime, that allows to study the two-excitation manifold of the JC ladder), and we found that the phenomenological dissipative JC model is mainly in the single-photon regime, while the driven QNM-JC model has multiphoton character.
Thus the quantum dissipation effects,
that appear in the quantized QNM model,
induce higher-order correlations.
This shows that the inter-mode coupling, coming from the microscopic QNM model, is crucial to include in the quantum master
equations.

\section{Acknowledgements}
We acknowledge support from the Deutsche Forschungsgemeinschaft (DFG) through SFB 951 Project B12 (Project number 182087777), Project BR1528/8-2 (Project number 177864488) and the 
Alexander von Humboldt Foundation through a Humboldt Research Award.
We also acknowledge funding from Queen's University,
the Canadian Foundation for Innovation, 
the Natural Sciences and Engineering Research Council of Canada, and CMC Microsystems for the provision of COMSOL Multiphysics.
This project has also received
funding  from  the  European  Unions  Horizon  2020
research and innovation program under Grant Agreement
No. 734690 (SONAR).
We thank 
Philip Tr\o st Kristensen,
Kurt Busch, and
Mohsen Kamandar Dezfouli for useful discussions.

\appendix

 \section{Input and output operators in the Markov approximation\label{app: InputOpApprox}}
In this appendix, we provide a more detailed derivation of the quantum Langevin equation in the Markov approximation, which is the basis for the QNM quantum master equation, Eq.~\eqref{eq: QNMMaster}, and the generalized input-output relations, Eq.~\eqref{eq:inout}. We start with the Heisenberg equation of motion of the QNM operator $a_\mu$ with respect to the Hamiltonian $H$ from Eqs.~\eqref{eq: Ha}-\eqref{eq: HI} and by using the expansion of the medium-assisted electric field in QNMs from Eq.~\eqref{eq:Esymm_multi},
\begin{equation}
\dot{a}_\mu=-i\int{\rm d}\mathbf{r}\int_0^\infty {\rm d}\omega~\omega \mathbf{L}_\mu(\mathbf{r},\omega)\cdot\mathbf{b}(\mathbf{r},\omega)-i\tilde{g}_\mu^{s*}\sigma^-,\label{eq: QLEApp0}
\end{equation}
where $\tilde{g}_\mu^s$ is TLS-QNM coupling constant in the symmetrized basis, given below Eq.~\eqref{eq:QLE}, and we have formally introduced $\mathbf{L}_{\mu}(\mathbf{r},\omega)$ via
\begin{align}
a_\mu &= \int_0^\infty {\rm d}\omega\int{\rm d}\mathbf{r}\mathbf{L}_{\mu}(\mathbf{r},\omega)\cdot\mathbf{b}(\mathbf{r},\omega),\label{eq: FormalQNMOp}
\end{align}
which reads explicitly (by comparing with Eq.~\eqref{eq: aDef}),
\begin{align}
    \mathbf{L}_{\mu}(\mathbf{r},\omega)&=\sum_{\eta}\left[\mathbf{S}^{-1/2}\right]_{\mu\eta}\sqrt{\frac{\omega}{2\pi\omega_\eta}}\frac{\sqrt{\omega\epsilon_I(\mathbf{r},\omega)}}{\tilde{\omega}_\eta-\omega}\tilde{\mathbf{f}}_{\eta}(\mathbf{r}),\label{eq: Lconst}
\end{align}
where $\tilde{\mathbf{f}}_{\eta}(\mathbf{r})$ is replaced by $\tilde{\mathbf{F}}_{\eta}(\mathbf{r},\omega)$ for positions $\mathbf{r}$ outside the resonator region. This can be viewed as a projection of the $\mathbf{b}(\mathbf{r},\omega)$ operators onto the QNM subspace.
Next, we define $\mathbf{c}(\mathbf{r},\omega)$ as the orthogonal complement to $a_\mu$ using
\begin{equation}
\mathbf{b}(\mathbf{r},\omega)=\sum_\mu\mathbf{L}_{\mu}^*(\mathbf{r},\omega)a_\mu + \mathbf{c}(\mathbf{r},\omega).\label{eq: Separation}
\end{equation}
This separation and resulting mapping of the full degrees of freedom onto a small subspace is commonly done for open quantum systems, as discussed in Ref.~\onlinecite{lambert2019modelling} or applied to a similar system in Ref.~\onlinecite{castellini2018quantum}.
Using Eq.~\eqref{eq: Separation} we can rewrite Eq.~\eqref{eq: QLEApp0} as
\begin{align}
\dot{a}_\mu=&-i\sum_{\mu'}\left[\int{\rm d}\mathbf{r}\int_0^\infty {\rm d}\omega~\omega \mathbf{L}_\mu(\mathbf{r},\omega)\cdot\mathbf{L}_{\mu'}^*(\mathbf{r},\omega)\right]a_{\mu'}\nonumber \\
&-i\tilde{g}_\mu^{s*}\sigma^--i \int{\rm d}\mathbf{r}\int_0^\infty {\rm d}\omega~\omega\mathbf{L}_{\mu}(\mathbf{r},\omega)\cdot\mathbf{c}(\mathbf{r},\omega).
\end{align}
Using Eq.~\eqref{eq: Lconst}, we can further write 
\begin{align}
    \dot{a}_\mu=&-i\sum_{\mu'}\left[\int{\rm d}\mathbf{r}\int_0^\infty {\rm d}\omega~\omega \mathbf{L}_\mu(\mathbf{r},\omega)\cdot\mathbf{L}_{\mu'}^*(\mathbf{r},\omega)\right]a_{\mu'} \nonumber\\
&-i\tilde{g}_\mu^{s*}\sigma^- -i \sum_{\eta}\left[\mathbf{S}^{-1/2}\right]_{\mu\eta}\int{\rm d}\mathbf{r}\int_0^\infty {\rm d}\omega\sqrt{\frac{\omega}{2\pi\omega_\eta}}\nonumber\\
&\times\frac{\omega-\tilde{\omega}_\eta+\tilde{\omega}_\eta}{\tilde{\omega}_\eta-\omega}\sqrt{\omega\epsilon_I(\mathbf{r},\omega)}\tilde{\mathbf{f}}_{\eta}(\mathbf{r})\cdot\mathbf{c}(\mathbf{r},\omega),
\end{align}
where we have added $-\tilde{\omega}_\eta+\tilde{\omega}_\eta$ in the numerator. Next, we separate the numerator into a term proportional to $\omega-\tilde{\omega}_\eta$ and a term proportional to $\tilde{\omega}_\eta$:
\begin{align}
    \dot{a}_\mu=&-i\sum_{\mu'}\left[\int{\rm d}\mathbf{r}\int_0^\infty {\rm d}\omega~\omega \mathbf{L}_\mu(\mathbf{r},\omega)\cdot\mathbf{L}_{\mu'}^*(\mathbf{r},\omega)\right]a_{\mu'}\nonumber \\
&-i\tilde{g}_\mu^{s*}\sigma^- -i \int{\rm d}\mathbf{r}\int_0^\infty {\rm d}\omega~\mathbf{g}_{\mu}(\mathbf{r},\omega)\cdot\mathbf{c}(\mathbf{r},\omega)\nonumber\\
&-i \sum_{\eta,\eta'}\left[\mathbf{S}^{-1/2}\right]_{\mu\eta}\tilde{\omega}_\eta\left[\mathbf{S}^{1/2}\right]_{\eta\eta'}\nonumber\\
&\times\int{\rm d}\mathbf{r}\int_0^\infty {\rm d}\omega\mathbf{L}_{\eta'}(\mathbf{r},\omega)\cdot\mathbf{c}(\mathbf{r},\omega),\label{eq: QLEApp2}
\end{align}
where
\begin{equation}
\mathbf{g}_{\mu}(\mathbf{r},\omega)=-\sum_{\eta}\left[\mathbf{S}^{-1/2}\right]_{\mu\eta}\sqrt{\frac{\omega}{2\pi\omega_\eta}}\sqrt{\omega\epsilon_I(\mathbf{r},\omega)}\tilde{\mathbf{f}}_{\eta}(\mathbf{r})\label{eq: gconst}.
\end{equation}
Using the orthogonality relation 
\begin{equation}
\int{\rm d}\mathbf{r}\int_0^\infty {\rm d}\omega~\mathbf{L}_{\mu}(\mathbf{r},\omega)\cdot\mathbf{c}(\mathbf{r},\omega)=0,
\end{equation}
which follows from Eq.~\eqref{eq: Separation} in combination with Eq.~\eqref{eq: FormalQNMOp} together with the orthonormality relation 
\begin{equation}
    \int{\rm d}\mathbf{r}\int_0^\infty {\rm d}\omega~ \mathbf{L}_\mu(\mathbf{r},\omega)\cdot\mathbf{L}_{\mu'}^*(\mathbf{r},\omega)=\delta_{\mu\mu'}
\end{equation}
for all $\mu,\mu'$, we obtain then
\begin{align}
\dot{a}_\mu=&-i\sum_{\mu'}\left[\int{\rm d}\mathbf{r}\int_0^\infty {\rm d}\omega~\omega \mathbf{L}_\mu(\mathbf{r},\omega)\cdot\mathbf{L}_{\mu'}^*(\mathbf{r},\omega)\right]a_{\mu'}\nonumber \\
&-i\tilde{g}_\mu^{s*}\sigma^- -i \int{\rm d}\mathbf{r}\int_0^\infty {\rm d}\omega~\mathbf{g}_{\mu}(\mathbf{r},\omega)\cdot\mathbf{c}(\mathbf{r},\omega)\label{eq: QLEPre}.
\end{align}
In addition, it can be shown, that the Hamiltonian from Eq.~(\eqref{eq: Ha}-\eqref{eq: HI}) can be recast into the form $H=H_{\rm sys}+H_{\rm sys-r}+H_{\rm r}$ with
\begin{align}
H_{\rm sys}=&\hbar\sum_{\mu,\mu'}\left[\int{\rm d}\mathbf{r}\int_0^\infty {\rm d}\omega~\omega \mathbf{L}_\mu(\mathbf{r},\omega)\cdot\mathbf{L}_{\mu'}^*(\mathbf{r},\omega)\right]a_{\mu}^\dagger a_{\mu'}\nonumber\\
&+H_a+H_{\rm L}+H_{\rm em-a},\\
H_{\rm sys-r}=& \hbar\sum_\mu\int_0^\infty{\rm d}\omega\int {\rm d}\mathbf{r}\mathbf{g}_{\mu}^*(\mathbf{r},\omega)\cdot\mathbf{c}^\dagger(\mathbf{r},\omega)a_\mu +{\rm H.a.},\\
H_{\rm r}=& \hbar\int_0^\infty {\rm d}\omega~\omega \int{\rm d}\mathbf{r}\mathbf{c}^\dagger(\mathbf{r},\omega)\cdot\mathbf{c}(\mathbf{r},\omega), 
\end{align}
where $H_{\rm L}$ and $H_{\rm em-a}$ are defined below Eq.~\eqref{eq:QLE}.

Next, we apply three approximations to connect to the Markovian quantum theory from Ref.~\onlinecite{Gardiner1} and the derivation of the QNM master equation from Ref.~\onlinecite{PhysRevLett.122.213901}:

($i$) We do a resonance approximation in the first term of Eq.~\eqref{eq: QLEPre}, 
\begin{align}
&\left[\int{\rm d}\mathbf{r}\int_0^\infty {\rm d}\omega~\omega \mathbf{L}_\mu(\mathbf{r},\omega)\cdot\mathbf{L}_{\mu'}^*(\mathbf{r},\omega)\right]\nonumber\\
&=\sum_{\eta,\eta'}\left[\mathbf{S}^{-1/2}\right]_{\mu\eta}\left[\int_0^\infty {\rm d}\omega~\omega S_{\eta\eta'}(\omega)\right]\left[\mathbf{S}^{-1/2}\right]_{\eta'\mu'},
\end{align}
where $S_{\eta\eta'}(\omega)$ is implicitly defined via $S_{\eta\eta'}=\int_0^\infty{\rm d}\omega S_{\eta\eta'}(\omega)$ from Eq.~\eqref{eq: Scommute}. Since $S_{\eta\eta'}(\omega)$ is dominated by the poles at $\omega=\tilde{\omega}_\eta$ and $\omega=\tilde{\omega}_{\eta'}^*$, we can (approximately) apply the residue theorem (after separating the poles via partial fraction) to get
\begin{align}
&\sum_{\eta,\eta'}\left[\mathbf{S}^{-1/2}\right]_{\mu\eta}\left[\int_0^\infty {\rm d}\omega~\omega S_{\eta\eta'}(\omega)\right]\left[\mathbf{S}^{-1/2}\right]_{\eta'\mu'}
    \nonumber\\
   &\approx \sum_{\eta,\eta'}\left[\mathbf{S}^{-1/2}\right]_{\mu\eta}\left[\frac{1}{2}(\tilde{\omega}_\eta+\tilde{\omega}_{\eta'}^* )S_{\eta\eta'}\right]\left[\mathbf{S}^{-1/2}\right]_{\eta'\mu'},
\end{align}
which is precisely $\chi^{(+)}_{\mu\mu'}$ from Eq.~\eqref{eq:QLE}.

($ii$) We approximate $\mathbf{c}(\mathbf{r},\omega)$ as bosonic operators, i.e., 
\begin{align}
[c_i(&\mathbf{r},\omega),c^\dagger_j(\mathbf{r}',\omega')]\nonumber\\
&=\delta_{ij}\delta(\mathbf{r}-\mathbf{r}')\delta(\omega-\omega')-\sum_\mu L_{\mu,i}^*(\mathbf{r},\omega)L_{\mu,j}(\mathbf{r}',\omega')\nonumber\\
&\approx \delta_{ij}\delta(\mathbf{r}-\mathbf{r}')\delta(\omega-\omega'). 
\end{align}
This leads to the time evolution,
\begin{align}
\mathbf{c}(\mathbf{r},\omega,t)=&\mathbf{c}(\mathbf{r},\omega,t_0)e^{-i\omega(t-t_0)}\nonumber\\
&-i\sum_\mu \mathbf{g}_{\mu}^*(\mathbf{r},\omega)\int_{t_0}^t{\rm d}t' e^{-i\omega(t-t')}a_\mu(t'),\label{eq: HeisnbergCOp}
\end{align}
with some fixed time $t_0<t$, i.e., the retarded solution. 
Inserting Eq.~\eqref{eq: HeisnbergCOp} into Eq.~\eqref{eq: QLEPre} leads to 
\begin{align}
\dot{a}_\mu=&-i\sum_{\mu'}\chi^{(+)}_{\mu\mu'}a_{\mu'} -i\tilde{g}_\mu^{s*}\sigma^-\nonumber\\
&-i \int{\rm d}\mathbf{r}\int_0^\infty {\rm d}\omega~\mathbf{g}_{\mu}(\mathbf{r},\omega)\cdot \mathbf{c}(\mathbf{r},\omega,t_0)e^{-i\omega(t-t_0)}\nonumber\\
&-\sum_{\mu'}\int{\rm d}\mathbf{r}\int_0^\infty {\rm d}\omega~\mathbf{g}_{\mu}(\mathbf{r},\omega)\cdot \mathbf{g}_{\mu'}^*(\mathbf{r},\omega)\nonumber\\
&\times\int_{t_0}^t{\rm d}t' e^{-i\omega(t-t')}a_{\mu'}(t').\label{eq: QLE1}
\end{align}

($iii$) We apply a Markov approximation. To do so, we look at
\begin{align}
&\int{\rm d}\mathbf{r}\mathbf{g}_{\mu}(\mathbf{r},\omega)\cdot \mathbf{g}_{\mu'}^*(\mathbf{r},\omega)\nonumber\\
=&\sum_{\eta,\eta'}\left[\mathbf{S}^{-1/2}\right]_{\mu\eta}\int{\rm d}\mathbf{r}\frac{\omega^2\epsilon_I(\mathbf{r},\omega)\tilde{\mathbf{f}}_{\eta}(\mathbf{r})\cdot\tilde{\mathbf{f}}_{\eta'}^*(\mathbf{r})}{2\pi\sqrt{\omega_\eta\omega_{\eta'}}}\left[\mathbf{S}^{-1/2}\right]_{\eta'\mu'}\nonumber\\
=&\frac{1}{2\pi}\sum_{\eta,\eta'}\left[\mathbf{S}^{-1/2}\right]_{\mu\eta}i(\tilde{\omega}_\eta-\tilde{\omega}_{\eta'}^*)\nonumber\\
&\times\left[\int{\rm d}\mathbf{r}\frac{\omega^2\epsilon_I(\mathbf{r},\omega)\tilde{\mathbf{f}}_{\eta}(\mathbf{r})\cdot\tilde{\mathbf{f}}_{\eta'}^*(\mathbf{r})}{i(\tilde{\omega}_\eta-\tilde{\omega}_{\eta'}^*)\sqrt{\omega_\eta\omega_{\eta'}}}\right]\left[\mathbf{S}^{-1/2}\right]_{\eta'\mu'}.
\end{align}
The relevant QNM frequencies in the system shall be enclosed by a small frequency interval $\Delta\omega$ (cf. App.~\ref{app: ReservoirPLusFreq} for discussion), which is usually the case in quantum optics, and is consistent with the rotating wave approximation from subsection~\ref{Subsec: GFquant}. Within a Markov approximation, we then pull $\int{\rm d}\mathbf{r}\mathbf{g}_{\mu}(\mathbf{r},\omega)\cdot \mathbf{g}_{\mu'}^*(\mathbf{r},\omega)$ out  of $\omega$-integral in Eq.~\eqref{eq: QLE1} and evaluate it within the small frequency interval $\Delta\omega$ to get
\begin{align}
\dot{a}_\mu\approx&-i\sum_{\mu'}\chi^{(+)}_{\mu\mu'}a_{\mu'}-i\tilde{g}_\mu^{s*}\sigma^- \nonumber\\
&-2\sum_{\mu'}\chi^{(-)}_{\mu\mu'}\frac{1}{2\pi}\int_0^\infty {\rm d}\omega\int_{t_0}^t{\rm d}t' e^{-i\omega(t-t')}a_{\mu'}(t')\nonumber\\
&-\sqrt{2}\sum_{\mu'}\left[\left(\boldsymbol{\chi}^{(-)}\right)^{1/2}\right]_{\mu\mu'}a_{\mu'}^{\rm in}\nonumber\\
=&-i\sum_{\mu'}\chi^{(+)}_{\mu\mu'}a_{\mu'} -i\tilde{g}_\mu^{s*}\sigma^--\sum_{\mu'}\chi^{(-)}_{\mu\mu'}a_{\mu'}\nonumber\\
&-\sqrt{2}\sum_{\mu'}\left[\left(\boldsymbol{\chi}^{(-)}\right)^{1/2}\right]_{\mu\mu'}a_{\mu'}^{\rm in},\label{eq: QLE2}
\end{align}
where
\begin{align}
    \chi^{(-)}_{\mu\mu'}&=\int{\rm d}\mathbf{r}\mathbf{g}_{\mu}(\mathbf{r},\omega)\cdot \mathbf{g}_{\mu'}^*(\mathbf{r},\omega)\big\vert_{\Delta\omega}\nonumber\\
    &\approx \sum_{\eta,\eta'}\left[\mathbf{S}^{-1/2}\right]_{\mu\eta}i(\tilde{\omega}_\eta-\tilde{\omega}_{\eta'}^*)S_{\eta\eta'}\left[\mathbf{S}^{-1/2}\right]_{\eta'\mu'},
\end{align}
which is consistent with the approximation in (i) and gives precisely the dissipation matrix from Eq.~\eqref{eq:QLE}, and
\begin{align}
a_\mu^{\rm in}=&\frac{i}{\sqrt{2}}\sum_{\mu'}\left[\left(\boldsymbol{\chi}^{(-)}\right)^{-1/2}\right]_{\mu\mu'} \nonumber\\
&\times\int{\rm d}\mathbf{r}\int_0^\infty {\rm d}\omega~\mathbf{g}_{\mu'}(\mathbf{r},\omega)\cdot \mathbf{c}(\mathbf{r},\omega,t_0)e^{-i\omega(t-t_0)}\label{eq: aInOpMarkov}
\end{align}
is the input operator with $t_0<t$ in the Markov approximation. Note, that in the second step of Eq.~\eqref{eq: QLE2}, we have used~\cite{Gardiner1} 
\begin{align}
    &\frac{1}{2\pi}\int_0^\infty {\rm d}\omega\int_{t_0}^t{\rm d}t' e^{-i\omega(t-t')}a_{\mu'}(t')\nonumber\\
    &\approx \int_{t_0}^t{\rm d}t'\delta(t-t')a_{\mu'}(t')=\frac{1}{2}a_{\mu'}(t),
\end{align}
where the approximation stems from extending the lower border of the $\omega$-integral from $0$ to $-\infty$, which is usually applied in quantum optics (cf. Ref.~\onlinecite{carmichael2009statistical}). 
Furthermore, the input operators $a_\mu^{\rm in}$ fulfil the commutation relation 
\begin{equation}
\left[a_\mu^{\rm in},a_{\mu'}^{\rm in\dagger}\right]\approx\delta_{\mu\mu'}\delta(t-t'),
\end{equation} 
which follows from the same Markov approximation as above. Eq.~\eqref{eq: QLE2} is now identical to the QLE in Eq.~\eqref{eq:QLE}, after evaluating the commutator $-i[a_\mu,H_{\rm sys}]/\hbar$ explicitly in Eq.~\eqref{eq:QLE}. 

To obtain the output operators, we choose a fixed time $t_1 > t$ in Eq.~\eqref{eq: HeisnbergCOp}, i.e., the advanced solution, to obtain the time-reversed QLE~\cite{Gardiner1}
\begin{align}
\dot{a}_\mu=&-i\sum_{\mu'}\chi^{(+)}_{\mu\mu'}a_{\mu'} -i\tilde{g}_\mu^{s*}\sigma^-+\sum_{\mu'}\chi^{(-)}_{\mu\mu'}a_{\mu'} \nonumber\\
&-\sqrt{2}\sum_{\mu'}\left[\left(\boldsymbol{\chi}^{(-)}\right)^{1/2}\right]_{\mu\mu'}a_{\mu'}^{\rm out},\label{eq: QLE2Rev}
\end{align}
where 
\begin{align}
    a_{\mu}^{\rm out}=&\frac{i}{\sqrt{2}}\sum_{\mu'}\left[\left(\boldsymbol{\chi}^{(-)}\right)^{-1/2}\right]_{\mu\mu'} \nonumber\\
    &\times\int{\rm d}\mathbf{r}\int_0^\infty {\rm d}\omega~\mathbf{g}_{\mu'}(\mathbf{r},\omega)\cdot \mathbf{c}(\mathbf{r},\omega,t_1)e^{-i\omega(t-t_1)}\label{eq: aOutOpMarkov}
\end{align}
are the output operators and the sign change in the third term of Eq.~\eqref{eq: QLE2Rev} is induced by change of time order between $t_0$ and $t_1$ with respect to $t$:
\begin{align}
    &\frac{1}{2\pi}\int_0^\infty {\rm d}\omega\int_{t_1}^t{\rm d}t' e^{-i\omega(t-t')}a_{\mu'}(t')\nonumber\\
    &\approx \int_{t_1}^t{\rm d}t'\delta(t-t')a_{\mu'}(t')=-\frac{1}{2}a_{\mu'}(t).
\end{align}
Combining Eq.~\eqref{eq: QLE2} and Eq.~\eqref{eq: QLE2Rev} then yields the input-output relations from Eq.~\eqref{eq:inout}. 
 
\section{Parameters and quasinormal mode
calculations for the hybrid cavity and two-level system\label{app: OG_Params}}

In this appendix, we briefly report on the numerical calculations of the hybrid structure in Fig.~\ref{fig: Scheme} and give more details on the QNM  and TLS parameters, as well as the calculation of the classical Purcell factor and $\beta$-factor.

To obtain the QNMs, a full three-dimensional Maxwell model was used to simulate the hybrid structure. Here, the dielectric constant of the plasmonic ellipsoid was modelled with the local Drude model
\begin{equation}
    \epsilon(\omega)=1-\frac{\omega_{\rm p}^2}{\omega(\omega+i\gamma_{\rm p})},
\end{equation}
with $\hbar\omega_{\rm p}=8.2934 \,{\rm eV}$ and $\hbar\gamma_{\rm p}=0.0928\,{\rm eV}$, while the PC beam is modelled via a constant refractive index $n_{\rm pc}=\sqrt{\epsilon_{\rm pc}}=2.04$. We note  that taking a constant permittivity $\epsilon\neq 1$ does not contradict with Kramers-Kronig relations here, since in the special case of a few-mode expansion, we restrict the $\omega$-integration in the quantum model to a finite interval (cf. App.~\ref{app: ReservoirPLusFreq} for details). Furthermore, the hybrid is embedded in a lossless background medium with $\epsilon_{\rm B}=1$. 
For a more detailed discussion on the QNM calculation of the hybrid, cf. Ref.~\onlinecite{ren2020near}. 
We note that non-local effects
can be included in the calculation of the
QNMs~\cite{KamandarDezfouli2017},
but are negligible for the
2-nm gap sizes studied here.

The two fundamental QNM (complex) eigenfrequencies of the hybrid structure are calculated as $\hbar\tilde{\omega}_{\rm pl}=1.6999 - 0.0479i~({\rm eV)}$ originating from the metallic ellipsoidal dimer, and $\hbar\tilde{\omega}_{\rm pc}=1.6052 - 0.0007i~({\rm eV)}$, originating from the PC beam. The dipole projected QNM eigenfunctions at the position of the $z$-polarized TLS (in the center of the plasmonic ellipsoid) are calculated as $\mathbf{n}_d\cdot \mathbf{\tilde{f}}_{{\rm pl}}(\mathbf{r}_{\rm a})=1.8002 \cdot 10^{12}~{\rm m}^{-3/2}  - 4.6917i \cdot 10^{10}~{\rm m}^{-3/2} $ and $\mathbf{n}_d\cdot \mathbf{\tilde{f}}_{\rm pc}(\mathbf{r}_{\rm a})=2.1079 \cdot 10^{11}~{\rm m}^{-3/2}  + 9.6228i \cdot 10^{10}~{\rm m}^{-3/2} $, where $\mathbf{n}_d=\mathbf{e}_z$. The elements of the intermode coupling matrix $\mathbf{S}$ are determined as $S_{\rm pl,pl}=0.068 + 0.894$, $S_{\rm pc,pc}=0.134 + 0.904$ and $S_{\rm pl,pc}=(-0.0021-0.0024i) + (-0.0042-0.0967i)$ with $S_{\rm pc,pl}=S_{\rm pl,pc}^*$, and where the first and second entry denotes the non-radiative and radiative part, respectively. Furthermore, we choose a dipole moment of $d_{\rm a}=10$ Debye, which is in the range of common values for quantum emitters in nano cavities~\cite{chikkaraddy2016single, gross2018near}. The QNM calculations and derivation of involving spatial integrals are performed with COMSOL~\cite{comsol}  and further details can be found in Ref.~\onlinecite{ren2020near}.

\section{Derivation of Eq.~\eqref{eq: IoutQNMfinal} and \eqref{eq: g2Final}\label{app: G2outDerivation}}
We start with the surface integral expression in the output intensity,
\begin{equation}
\bar{I}^{\rm out} =\oint_\mathcal{S} {\rm d}A_\mathbf{s}\langle \hat{E}_{{\rm out},i}^{(-)}(\mathbf{s})\hat{E}_{{\rm out},i}^{(+)}(\mathbf{s})\rangle\label{eq: barI1}.
\end{equation}
Inserting the expression from Eq.~\eqref{eq: Eout3} into Eq.~\eqref{eq: barI1} and using the fact, that all normal-ordered expectation values involving $A_\mu^{\rm in}$ vanish, we obtain
\begin{equation}
  \bar{I}^{\rm out}=\frac{\hbar }{2\epsilon_0n_{\rm B} c} \sum_{\mu,\eta} L_{\mu\eta} \langle A_\mu^\dagger A_\eta\rangle,
\end{equation}
with 
\begin{equation}
    L_{\mu\eta}=2n_{\rm B }c\sqrt{\omega_\mu\omega_\eta}\sqrt{\Gamma_\mu\Gamma_\eta}\oint_{\mathcal{S}}\mathbf{Z}_{\mu}^{\rm sU*}(\mathbf{s})\cdot \mathbf{Z}_{\eta}^{\rm sU}(\mathbf{s}){\rm d}A_\mathbf{s}.
\end{equation}
Using the definition of $\mathbf{Z}_{\eta}^{\rm sU}(\mathbf{s})$ from Eq.~\eqref{eq: FregDiagU} in combination with Eq.~\eqref{eq: ApproxFreg}, we arrive at 
\begin{align}
    L_{\mu\eta}=&n_{\rm B }c\sum_{\eta',\eta''}\sum_{\mu',\mu''}\sqrt{\omega_\eta'\omega_\mu'}\left(\mathbf{S}^{1/2}\right)_{\mu''\mu'}U^{(-)*}_{\mu\mu''}\left(\mathbf{S}^{1/2}\right)_{\eta'\eta''}\nonumber\\
     &\times U^{(-)}_{\eta\eta''}\oint_{\mathcal{S}}\tilde{\mathbf{F}}_{\mu''}^*(\mathbf{s},\omega_{\mu''})\cdot \tilde{\mathbf{F}}_{\eta''}(\mathbf{s},\omega_{\eta''}){\rm d}A_\mathbf{s}\label{eq: Lpre}.
\end{align}
Looking at the form of $S^{\rm rad}_{\mu\eta}$ in Eq.~\eqref{eq: Srad}, and 
recognizing  that the frequency dependence is dominated by Lorentz functions at the QNM frequencies and assuming that the regularized QNMs $\tilde{\mathbf{F}}_{\mu}^*(\mathbf{s},\omega_{\mu})$ are constant over the frequency regime that includes the relevant QNMs, we can write   
\begin{align}
    n_{\rm B }c\oint_{\mathcal{S}}\tilde{\mathbf{F}}_{\mu}^*(\mathbf{s},\omega_{\mu})\cdot \tilde{\mathbf{F}}_{\eta}(\mathbf{s},\omega_{\eta}){\rm d}A_\mathbf{s}\approx \frac{2i(\tilde{\omega}_{\eta}-\tilde{\omega}_{\mu}^*)}{\omega_\mu + \omega_\eta}S_{\eta\mu}^{\rm rad}.
\end{align}
Inserting this approximate form into Eq.~\eqref{eq: Lpre} yields the final form of Eq.~\eqref{eq: Lfinal}. 

Finally, we also define 
\begin{equation}
    I^{\rm out}=\frac{2\epsilon_0n_{\rm B} c}{\hbar}\bar{I}^{\rm out},
\end{equation}
to arrive at Eq.~\eqref{eq: IoutQNMfinal}. The derivation of Eq.~\eqref{eq: g2Final} can be performed in the same manner upon using the above arguments.

 \section{Discussion on the emitter-photon coupling regimes\label{app: CouplingRegime}}
\begin{figure}[!h]
	\centering
	\includegraphics[trim={0.20cm 0 0.15cm 0},clip]{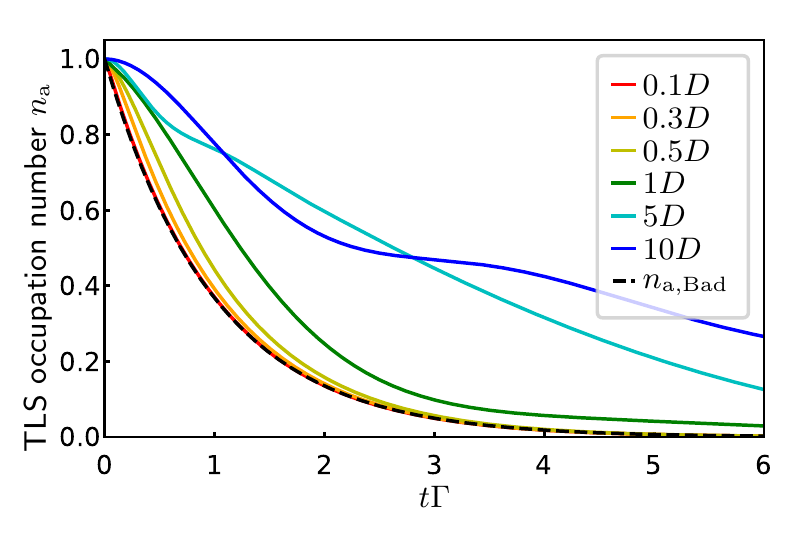}
	\caption{Excited state occupation $n_{\rm a}=\langle \sigma^+\sigma^-\rangle$ over time for different dipole strengths in units of Debye ($D$), using the quantized QNM model for the hybrid structure with initial occupation $n_{\rm a}(t=0)=1$ and without external pump. The time axis is normalized to the cavity-enhanced spontaneous emission rate $\Gamma$ from Eq.~\eqref{eq: GammaEnhQNM} scaled with the respective dipole strength. $n_{\rm a,Bad}$ (black dashed line) reflects the exponential decay in the bad-cavity limit, i.e., $n_{\rm a,Bad}(t)=\exp[-(\Gamma+\gamma_{\rm SE}) t]$. Note, that for the given dipole strengths, $\Gamma\gg \gamma_{\rm SE}$, which implies the same bad cavity limit result $n_{\rm a,Bad}(t)\approx\exp[-\Gamma t]$ for all cases on the scaled time axis.
	}\label{fig: TLSDynamics}
\end{figure}

In this appendix, we discuss the two different light-matter coupling regimes, that were studies in Section~\ref{Sec: Applications}. To characterize the different coupling regimes, the temporal behaviour of the initially excited TLS (without external pump) is shown in Fig.~\ref{fig: TLSDynamics} for different dipole strengths.

For the Purcell factor and $\beta$-factor calculations, the dipole strength was set to $d\leq 0.1D$. As one can see in Fig.~\ref{fig: TLSDynamics} (red curve), in this regime, the TLS decays exponentially, and the temporal behaviour is practically identical to the bad cavity limit (black dashed line in Fig.~\ref{fig: TLSDynamics}), which clearly indicates the weak light-exciton coupling regime. For the probability and output correlation function simulations, the dipole strength was set to $d=10D$, leading to the QNM-TLS coupling constant as depicted in Tab.~\ref{Tab: Params}. For this choice, there is damped vacuum Rabi oscillations in the time dynamics of the TLS, as one can see in Fig.~\ref{fig: TLSDynamics} (dark blue curve), which clearly indicates the strong light-exciton coupling regime. Additionally, examples for the intermediate regime are shown ($0.3D-1D$), where the temporal behaviour of the TLS starts to deviate from the bad cavity limit results, but is still dominated by the exponential decay. We emphasize, that our quantized QNM model in its current form can in general be applied to light-matter coupling regimes, where the rotating wave-approximation is valid, which includes the strong and lower coupling regimes.

\section{Discussion on excitation regimes\label{app: ExcitationRegime}}
\begin{figure}[!h]
	\centering
	\includegraphics[trim={0.20cm 0 0.15cm 0},clip]{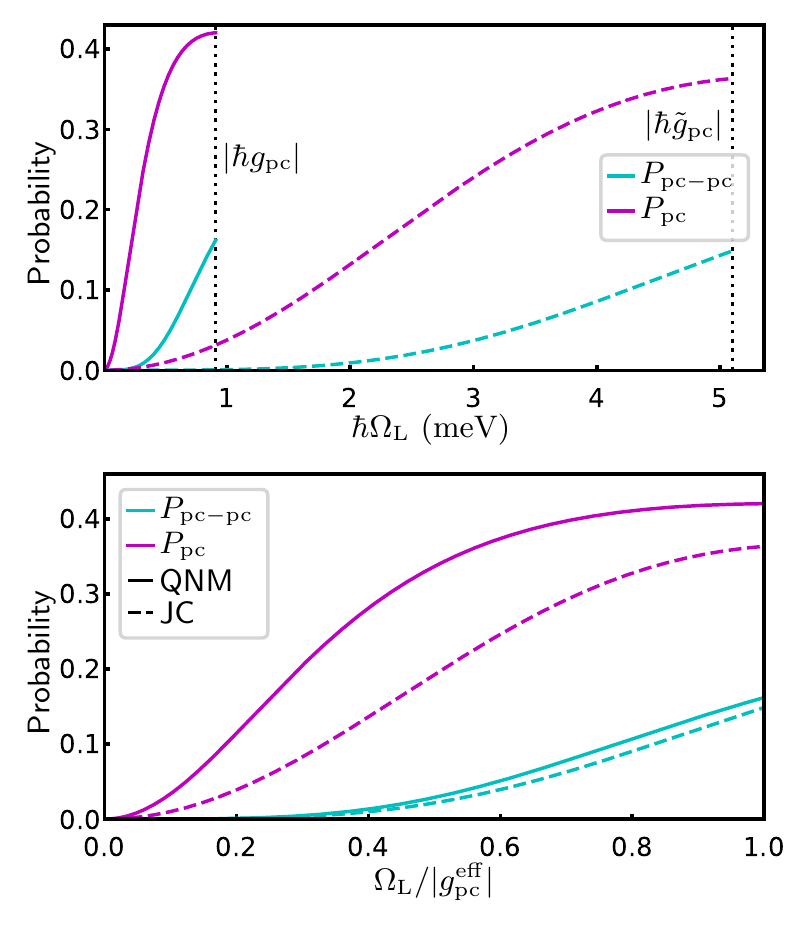}
	\caption{The 1- and 2-excitation probability of the PC-like eigenstate at the laser frequency of maximum probability peak for the QNM-JC (solid) and phenomenological dissipative JC model (dashed) over Rabi frequency $\Omega_{\rm L}$. Lower plot shows the same result as upper plot with scaled $\Omega_{\rm L}$ axis with $g_{\rm pc}^{\rm eff}=\tilde{g}_{\rm pc}$ for the phenomenological dissipative JC model and $g_{\rm pc}^{\rm eff}=g_{\rm pc}$ for the QNM-JC model. $\Omega_{\rm L}=g_{\rm pc}^{\rm eff}$ signifies the intermediate excitation regime. Both models have different intermediate excitation regimes, since $|\tilde{g}_{\rm pc}|/|g_{\rm pc}|\approx 5$. 
	}\label{fig: ProbOverRabi}
\end{figure}

In Section~\ref{Sec: Applications}, we compared the QNM-JC model with a phenomenological dissipative JC model, using the same external laser Rabi frequency $\Omega_{\rm L}\sim g_{\rm pc}$, where $g_{\rm pc}$ is the coupling constant between the PC mode and the TLS (cf. Tab.~\ref{Tab: Params}). Assuming that the PC-mode is the main coupling part for TLS frequencies near the PC-mode frequency,  $\Omega\sim |g_{\rm pc}|\approx 8\Gamma_{\rm pc}$ indicates the intermediate excitation regime. 
However, we emphasize, that the hybrid
system described by the phenomenological dissipative JC model has a larger TLS-PC coupling compared to the QNM-JC model, i.e., $|\tilde{g}_{\rm pc}|\approx 5|g_{\rm pc}|$. Therefore the excitation regimes are slightly different. Thus, another possibility to compare both models would be to choose a Rabi frequency, that scales with the different TLS-PC coupling constants, i.e, choosing a different Rabi frequency for the QNM-JC model and the phenomenological dissipative JC model. 
To discuss the different ways of comparison, we show the peaks of the eigenstate probabilities $P_{\rm pc}$, $P_{\rm pc-pc}$ (discussed in subsection~\ref{subsec: sysprop}) as function of $\Omega_{\rm L}$ in Fig.~\ref{fig: ProbOverRabi}. In Fig.~\ref{fig: ProbOverRabi} (top), we show results for the treatment of the Rabi frequency, which is used in the main part, i.e. we choose the same external laser strength for both models. In this case, the same hybrid response completely different to an external laser in both models. In contrast, in  Fig.~\ref{fig: ProbOverRabi} (bottom), a treatment, where the Rabi frequency scales with the different TLS-PC coupling constants is shown: Here, the probabilities are quiet similar, but a significant quantitative difference is still present.

 \section{Discussion on the treatment of the frequency integrals\label{app: ReservoirPLusFreq}}

 Here we discuss the treatment of the frequency integrals in the QNM quantization. Using the QNM Green function from Eq.~\eqref{eq: GreenQNM} together with the regularized QNMs (Eq.~\eqref{eq: DysonF} or Eq.~\eqref{eq: NF2FF_Z}), 
 the positive-rotating part of 
 the total electric field operator from Eq.~\eqref{eq: SolE} reads  
\begin{equation}
    \hat{\mathbf{E}}^{(+)}(\mathbf{r})=\int_0^\infty{\rm d}\omega\int {\rm d}\mathbf{r}\frac{i}{\epsilon_0\omega}\mathbf{G}_{\rm QNM}(\mathbf{r},\mathbf{r}',\omega)\cdot\hat{\mathbf{j}}_{\rm N} (\mathbf{r}',\omega),
\end{equation}
where $\mathbf{G}_{\rm QNM}(\mathbf{r},\mathbf{r}',\omega)$ has the form from Eq.~\eqref{eq: GreenQNM}, and $\tilde{\mathbf{f}}_\mu$ is replaced by the regularized QNM $\tilde{\mathbf{F}}_\mu$ for spatial positions outside the resonator region.
We next separate the frequency integral in $\hat{\mathbf{E}}^{(+)}_{\rm QNM}(\mathbf{r})$ into two integrals, one over a small frequency interval $\Delta\omega$ of interest (e.g., the optical regime), where few QNMs dominate, and the other over the complementary interval $\mathbb{R}^+/\Delta\omega$. We thus arrive at the formal separation
\begin{equation}
    \hat{\mathbf{E}}^{(+)}(\mathbf{r})=\hat{\mathbf{E}}^{(+)}(\mathbf{r})\big\vert_{\Delta\omega} + \hat{\mathbf{E}}^{(+)}(\mathbf{r})\big\vert_{\mathbb{R}^+/\Delta\omega},
\end{equation}
 where in both contributions, the full sum of QNMs still appears. However, we can approximate $\hat{\mathbf{E}}_{\rm QNM}^{(+)}(\mathbf{r})\big\vert_{\Delta\omega}$, by only using a subset $D_{\Delta\omega}\in \mathbb{Z}$ of QNMs, which are the dominant contributions in $\Delta\omega$, for the expansion. The resulting error of this approximation can be quantified with the overlap of $A_{\mu}(\omega)$, $\mu\in\mathbb{Z}/D_{\Delta\omega}$, into the frequency interval $\Delta\omega$.

\end{document}